\documentclass[12pt]{article}

\usepackage{graphicx}
\usepackage{amsmath}
\usepackage{amssymb}
\usepackage{hyperref}
\usepackage{cite}
\usepackage{times}
\usepackage{fullpage}
\usepackage{ifthen}
\usepackage{placeins}
\usepackage{epsfig}
\usepackage{deluxetable}
\usepackage{threeparttable}
\usepackage{xcolor}
\usepackage{lineno}
\usepackage{titling}

\usepackage{comment}
\usepackage{multibib}
\newcites{main}{References}
\newcites{method}{Methods References}

\def\hst        {{\em HST}\/}

\def\chandra    {{\em Chandra}\/}

\def\xmm        {{\em XMM-Newton}\/}

\def\ga         {{ESO~137-001}\/}
\def\gb         {{ESO~137-002}\/}

\def\gd         {{NGC~4848}\/}
\def\gv         {{NGC~4569}\/}

\newcommand{\hi}{H\,{\sc i}}
\newcommand{\hii}{H\,{\sc ii}}

\newcommand{\RNum}[1]{\uppercase\expandafter{\romannumeral #1\relax}}
\newcommand{\NII}{\mbox{[N\,\textsc{ii}]}}

\newenvironment{addendum}{%
    \setlength{\parindent}{0in}%
    \small%
    \begin{list}{Acknowledgements}{%
        \setlength{\leftmargin}{0in}%
        \setlength{\listparindent}{0in}%
        \setlength{\labelsep}{0em}%
        \setlength{\labelwidth}{0in}%
        \setlength{\itemsep}{12pt}%
        }
    }
    {\end{list}\normalsize}

\newcommand*{\addendumlabel}[1]{\textbf{#1}\hspace{1em}}

\newenvironment{methods}{%
    \section*{Methods}%
    \setlength{\parskip}{6pt}%
    }{}

\title{A universal correlation between warm and hot gas in the stripped tails of cluster galaxies}

\author{Ming Sun$^{1\star}$, 
Chong Ge$^{1\star}$, 
Rongxin Luo$^{1\star}$, 
Masafumi Yagi$^{2}$, \\
Pavel J\'{a}chym$^{3}$,
Alessandro Boselli$^{4}$,
Matteo Fossati$^{5,6}$,
Paul E.J. Nulsen$^{7,8}$, \\
Michitoshi Yoshida$^{9}$,
Giuseppe Gavazzi$^{5}$
}

\begin{document}

\maketitle

\newenvironment{affiliations}{%
    \setcounter{enumi}{1}%
    \setlength{\parindent}{0in}%
    \slshape\sloppy%
    \begin{list}{\upshape$^{\arabic{enumi}}$}{%
        \usecounter{enumi}%
        \setlength{\leftmargin}{0in}%
        \setlength{\topsep}{0in}%
        \setlength{\labelsep}{0in}%
        \setlength{\labelwidth}{0in}%
        \setlength{\listparindent}{0in}%
        \setlength{\itemsep}{0ex}%
        \setlength{\parsep}{0in}%
        }
    }{\end{list}\par\vspace{12pt}}

\begin{affiliations}
 \item Department of Physics and Astronomy, University of Alabama in Huntsville, Huntsville, AL 35899, USA
 \item National Astronomical Observatory of Japan, 2-21-1, Osawa, Mitaka, Tokyo, 181-8588, Japan
 \item Astronomical Institute of the Czech Academy of Sciences, Bo\v{c}n\'{i} II 1401, 141 00, Prague, Czech Republic
  \item Aix Marseille Universit\'{e}, CNRS, LAM (Laboratoire d’ Astro physique de Marseille) UMR 7326, 13388, Marseille, France
 \item Dipartimento di Fisica G. Occhialini, Universit\`{a} degli Studi di Milano Bicocca, Piazza della Scienza 3, I-20126 Milano, Italy
 \item INAF-Osservatorio Astronomico di Brera, via Brera 28, I-20121 Milano, Italy
 \item Center for Astrophysics \textbar{} Harvard \& Smithsonian, Cambridge, MA 02138, USA
 \item ICRAR, University of Western Australia, 35 Stirling Hwy, Crawley, WA 6009, Australia
 \item Subaru Telescope, National Astronomical Observatory of Japan, 650 North A'ohoku Place, Hilo, HI 96720, USA
\end{affiliations}

\renewenvironment{abstract}{%
    \setlength{\parindent}{0in}%
    \setlength{\parskip}{0in}%
    \bfseries%
    }{\par\vspace{0pt}}
\begin{abstract}

The impact of ram pressure stripping
on galaxy evolution is well known (e.g., \citemain{Boselli06}). Recent multi-wavelength data have revealed many examples of galaxies undergoing stripping, often accompanied with multi-phase tails (\citemain{Gavazzi01,Sun06,Chung07,Yagi07,Yoshida08,Yagi10,Smith10,Sivanandam10,Jachym14,Boselli16,Jachym17,Poggianti17}).
As energy transfer in the multi-phase medium is an outstanding question in astrophysics, 
galaxies in stripping are great objects to study.
Despite the recent burst of observational evidence, the relationship between gas in different phases in the tails is poorly known. Here we report a strong linear correlation between the X-ray surface brightness
and the H$\alpha$ surface brightness
of the diffuse gas in the stripped tails at $\sim$ 10 - 40 kpc scales,
with a slope of $\sim$ 3.5.
This discovery provides evidence for the mixing of the stripped interstellar medium with the hot intracluster medium as the origin of the multi-phase tails.
The established relation in stripped tails, also in comparison with the likely related correlations in similar environments like galactic winds and X-ray cool cores,
provides an important test for models of energy transfer in the multi-phase gas.
It also indicates the importance of the H$\alpha$ data to study clumping and turbulence in the intracluster medium.

\end{abstract}

We have constructed the largest sample of ram pressure stripping (RPS) galaxies in nearby clusters with both deep H$\alpha$ data (tracing warm gas with $T \sim 10^{4}$ K) from narrow-band imaging or {\em MUSE} integral field spectrograph (IFS) and deep X-ray data (tracing hot gas with $T \sim 10^{7}$ K) from \chandra\ or \xmm{} (Table~\ref{table:galaxy1} and Supplementary Tables 1-2).
All 17 galaxies have H$\alpha$ tails detected from previous works and 12 of them have X-ray tails detected, including 6 new ones presented in this work.
The tails are divided into 42 regions 
with sizes from 6.3 kpc$^{2}$
to 2.1$\times10^{3}$ kpc$^{2}$, all beyond $D_{25}$ (the isophotal level of 25 mag/arcsec$^{2}$ in the $B$-band) of the galaxy.
The X-ray surface brightness (SB) in these regions, with point sources excluded, is measured. If undetected, 5-$\sigma$ upper limits for the X-ray emission are derived (for eight regions in seven tails).
The X-ray bolometric fluxes and luminosities are also derived.
Similarly, the H$\alpha$ SB in these regions, with background sources and \hii\ regions excluded, is also measured. As shown in Fig.~\ref{fig:relation1}, there is a strong correlation between the X-ray SB and the H$\alpha$ SB in the tail regions. 
The correlation is well described by a simple linear relation of SB$_{\rm X}$/SB$_{\rm H\alpha} = 3.48 \pm 0.25$ (as also shown in the right panel of Fig.~\ref{fig:relation1}) and the intrinsic scatter of the correlation is small at $\sim$ 10\%.
Our simple linear relation is also different from the previous correlation derived from one galaxy \citemain{2019ApJ...887..155P}.

We can also examine individual H$\alpha$/X-ray tails. \ga\ hosts the brightest H$\alpha$/X-ray tail \citemain{Sun10,Fumagalli14}. We present a new H$\alpha$ image of \ga, with new {\em MUSE} data to achieve complete coverage of its X-ray tail. As shown in Fig.~\ref{fig:eso137001} and Supplementary Figures 1-2, the general positional correlation between H$\alpha$ and X-rays is very good. Both show double tails extending to at least 80 kpc from the galaxy.
We are able to study the H$\alpha$ --- X-ray correlation in 14 tail regions of \ga.
As is the case for every RPS tail, the study is limited by the X-ray data. 
Beyond $\sim$ 15 kpc from the nucleus, the {\em MUSE} exposure is only 15 - 35 minutes
per field, but sufficiently deep to probe the H$\alpha$ emission at $\sim$ kpc scales. By contrast, the \chandra\ exposure is 38.4 hours that covers the whole tail region and the median area of tail regions is 136 kpc$^{2}$. 
On the other hand, it is unclear whether such a correlation still holds at $\sim$ kpc scales, given the observed scatter at $\sim$ 10 - 40 kpc scales.
One upper limit for \ga\ is around some H$\alpha$ diffuse emission just beyond the primary X-ray tail (Fig.~\ref{fig:eso137001}). 
Further X-ray extension beyond what is shown in Fig.~\ref{fig:eso137001} is indeed suggested from the \chandra\ temperature map \citemain{Sun10}.
We also show the X-ray and H$\alpha$ images of other RPS tails in the Supplementary Figures 3-13. 
Generally good correlation between H$\alpha$ and X-ray can be seen but the data quality (almost always for X-rays) does not allow for more detailed studies at smaller spatial scales. The X-ray properties of these tails are presented in the Supplementary Table 3. 

We examined the correlation at similar region sizes of 300 - 440 kpc$^{2}$ and the results are the same (Supplementary Fig.~\ref{fig:relation1large}).
As the model with a constant X-ray/H$\alpha$ ratio describes the data well, we also attempted to examine whether this constant depends on the cluster. The derived ratios are 
3.1$\pm$1.0, 2.5$\pm$0.5, 2.8$\pm$1.1, 3.7$\pm$0.4 and 4.4$\pm$0.7 for Virgo, A2626, A1367, A3627 and Coma respectively. Thus, while the current data may suggest a small cluster-to-cluster variation, more data are required to examine the trend with cluster mass or pressure of the intracluster medium (ICM).
One factor to contribute to the scatter is the remaining \hii{} regions, or recent star formation (SF). While the flux fraction of \hii{} regions in our tail regions is typically very small (Supplementary Fig.~\ref{fig:HIIfraction}), the X-ray/H$\alpha$ ratio from SF is typically much smaller from our studies of galaxy regions (Supplementary Fig.~\ref{fig:relation2g2}).
We also examined whether the X-ray/H$\alpha$ ratio in the tail changes with the projected distance from the galaxy (Supplementary Fig.~\ref{fig:ratio_dist}).
The ratio likely increases at $>$ 60 kpc from the nucleus for one tail with the best X-ray data, but the results for other tails remain inconclusive.
If mixing between the stripped interstellar medium (ISM) and ICM eventually depletes the stripped cold gas, the depletion is not observed from the ratio between the soft X-ray gas and the warm gas within $\sim$ 60 kpc from the galaxy. However, it remains to see how the ratio of the hot (or warm) gas to the cold gas changes with the distance to the galaxy in stripped tails, ideally with the \hi\ and CO data.
We also examined the correlation between the X-ray/H$\alpha$ ratio and the distance of the galaxy from the cluster center, scaled by the virial radius or not, as well as the relative velocity of the galaxy to that of the cluster, scaled by the radial velocity dispersion of the cluster. No correlation was found.

Our results suggest that the formation of the X-ray tail is tightly related to the formation of the H$\alpha$ tail, which in turn indicates a strong correlation between the hot gas and the warm gas. This is also strong evidence for mixing between the stripped ISM and the hot ICM, at spatial scales smaller than $\sim$ 10 kpc, as good correlation has been established above this scale from the current data.
The emissivity of both phases, all proportional to density squared, may have a similar dependence on the ambient pressure. We attempt a simple model to study this ratio
and the results hinge on the detailed multi-phase gas distribution in small scales of the mixing layer (Supplementary Information).
The multi-phase stripped tails have not been well studied in simulations and the correlation between different phases has not been examined in detail. \citemain{Tonnesen11} gave the total X-ray and H$\alpha$ luminosities for stripped tails in three different simulation runs. The estimated X-ray/H$\alpha$ ratio is 1.3 - 4.6 in these three runs, with the lowest ratio coming from the run with the lowest ambient pressure.
More simulations are needed to better explore the H$\alpha$ - X-ray correlation at different spatial scales and compare them to the observed relation.
The evolution of this X-ray/H$\alpha$ ratio is another interesting question to be addressed by simulations and observations in the future.

Stripped tails have become another kind of unique environment where a multi-phase medium and SF are present, similar to multiphase galactic winds (e.g., \citemain{Strickland02}) and X-ray cool cores surrounding the brightest central galaxies in galaxy groups and clusters (e.g., \citemain{Fabian03}). Thus, it is natural to ask whether a similar relation between X-ray and H$\alpha$ exists in those two environments. However, such correlations in X-ray cool cores and galactic winds have not been studied in detail (Supplementary Information).
The existing works on galactic winds of nearby starbursts suggest a tight H$\alpha$/X-ray correlation with similar ratios to those found in stripped tails, while the X-ray-to-H$\alpha$ ratios in X-ray cool cores are typically smaller.
Future sample studies can explore the H$\alpha$ --- X-ray correlation in galactic winds and X-ray cool cores better.

In the stripped tails, the diffuse H$\alpha$ emission (excluding emission related to \hii\ regions or SF) likely originates from the turbulent mixing layer between the cold phase (molecular gas or/and atomic gas) and the hot ICM. Hydrodynamic instabilities and turbulence increase the contact surface between two phases and enhance mixing. We can consider the ratio between the turbulence eddy turnover timescale and the cooling timescale, $t_{\rm eddy} / t_{\rm cooling}$ --- the $C$-ratio (e.g., \citemain{Gaspari18}) or the Damk\"{o}hler number (e.g., \citemain{Tan21}). This ratio may vary with locations in the multi-phase medium in stripped tails. In the ICM around the stripped tails, $t_{\rm cooling}$ is almost always longer than 10 Gyr. $t_{\rm eddy}$ is 0.1 Gyr for a turbulence velocity of 100 km/s at a spatial scale of 10 kpc. Thus, the ratio should be much less than 1 for efficient mixing. When moving towards the stripped cold ISM, gas temperatures after mixing are lower and cooling can be much stronger to increase the $t_{\rm eddy} / t_{\rm cooling}$ ratio over 1, where inhomogeneous cooling results in a multiphase medium. For the $T \sim 10^{7}$ K gas in the stripped tails, $t_{\rm cooling}$ is typically 0.3 - 3 Gyr. $t_{\rm cooling}$ should be much smaller for $T \sim 10^{5-6}$ K gas.
With an analogy to galactic winds, the stripped cold ISM clouds can grow via mixing under certain conditions
(e.g., \citemain{Gronke18,Fielding20}), but can also fragment and even ``shatter'' in the surrounding ICM, likely forming a complex distribution of small clumps resembling droplets (e.g., \citemain{Gronke20}).
Magnetic field may also play an important role in mixing (e.g., \citemain{Sparre20}), as its existence is also implied by narrow H$\alpha$ filaments observed in many tails of our sample.

Besides the H$\alpha$ - X-ray (or warm gas - hot gas) correlation shown here, is there a tight correlation between other phases in stripped tails? 
The dominant mass component in stripped tails, at least at the early stage of stripping, is likely the molecular gas \citemain{Jachym14,Jachym17}. 
The cold gas in stripped tails, traced by \hi\ and CO, should eventually be evaporated or cool to form stars, resulting in an increasing hot (or warm) gas to cold gas ratio.
\citemain{Jachym17} explored the H$\alpha$ - CO (or warm gas - cold gas) correlation but more CO detections from the stripped tails are required to explore the correlation better. Moreover, while compact molecular clouds have been revealed in the best-studied RPS tail of \ga\ \citemain{Jachym19} (also shown in Fig.~\ref{fig:eso137001}), the same study also suggests most molecular gas is diffuse in the tail.
Future studies may need to separate the compact/diffuse molecular components in the correlation studies.

Evaporating cold gas stripped from galaxies also contributes to the ICM clumping.
The clumpiness of the ICM has been studied with X-ray observations, directly from e.g., SB fluctuation \citemain{Churazov12}, or indirectly from e.g., cluster outskirts and scaling relations \citemain{Nagai11}.
The ICM clumping can bias the measured X-ray properties of the cluster, e.g., density, gas mass and pressure, which can further bias the resulting mass of the cluster \citemain{Nagai11,2013MNRAS.429..799V}.
Characterization of ICM clumping is important for current and next generation surveys in the X-ray and millimeter via the Sunyaev-Zel'dovich effect, as well as using clusters as precise cosmology probes.
An important source of the ICM clumps is the stripped ISM, either by ram pressure or tidal force. When the stripped ISM evaporates in the ICM, it induces inhomogeneities. The tight correlation revealed by our studies suggests that at least some ICM clumps can be probed and even predicted by sensitive, wide-field H$\alpha$ surveys (e.g., \citemain{Boselli18}). 
This discovery opens a new window to use H$\alpha$ to trace hot gas structure in the ICM, including both the density fluctuations and the kinematic substructure.

\begin{comment}

\end{comment}

\begin{addendum}
\item[Correspondence] Correspondence and requests for materials should be addressed to Ming Sun~(ms0071@uah.edu), Chong Ge~(gc0034@uah.edu), Rongxin Luo~(rl0055@uah.edu).

\item[Acknowledgements]
M.S. thanks Andy Fabian, Yuan Li, Stephanie Tonnesen and Daniel Wang for helpful discussions. We thank Tim Edge and Sunil Laudari for work on the FIR data and the {\em HST} data. We also thank referees for helpful comments. Support for this work was provided by the National Aeronautics and Space Administration through \chandra\ Award Number GO6-17127X and GO6-17111X issued by the \chandra\ X-ray Center, which is operated by the Smithsonian Astrophysical Observatory for and on behalf of the National Aeronautics Space Administration under contract NAS8-03060.
Support for this work was also provided by the NASA grants 80NSSC19K0953 and the NSF grant 1714764.
P.J. acknowledges support from the project EU-ARC.CZ (LM2018106) of the Ministry of Education, Youth and Sports of the Czech Republic.
M.F. acknowledges support from the European Research Council (ERC) (grant agreement No 757535).
This research has made use of data obtained from the \chandra\ Data Archive and the \chandra\ Source Catalog, and software provided by the \chandra\ X-ray Center (CXC) in the application packages CIAO. This research is also based on observations obtained with \xmm, an ESA science mission with instruments and contributions directly funded by ESA Member States and NASA.  
This research is also based on observations collected at the European Southern Observatory under ESO programmes 60.A-9349(A), 60.A-9100(G), 095.A-0512(A), 096.B-0019(A), 098.B-0020(A), 0103.A-0684(A) and 0104.A-0226(A).
This research is based in part on data collected at Subaru Telescope, which is operated by the National Astronomical Observatory of Japan. We are honored and grateful for the opportunity of observing the Universe from Maunakea, which has the cultural, historical and natural significance in Hawaii.

\item[Author contributions] 
M.S. initiated the research, led the \chandra\ and \xmm\ proposals, and the {\em MUSE} proposals on \ga, analyzed the \chandra\ data, assisted the {\em MUSE} data analysis, and wrote the manuscript.
C.G. analyzed the \xmm\ data and R.L. analyzed the {\em MUSE} data of \ga\ and D100. Both contributed to the writing of the manuscript. P.J. is the PI of the {\em MUSE} proposal on the Coma galaxies. G.G. is the PI of the {\em MUSE} proposal on the A1367 galaxies. M.F. analyzed the {\em MUSE} data of A1367 galaxies. M.Y., M.F., G.G., A.B. and M.Y. provided the narrow-band H$\alpha$ imaging data. All authors contribute to the discussion and interpretation of the results.

\item[Competing interests] The authors declare that they have no
competing financial interests.
\end{addendum}

\newpage

\begin{figure*}[!htbp]
\begin{center}
\includegraphics[angle=0,width=1.0\columnwidth]{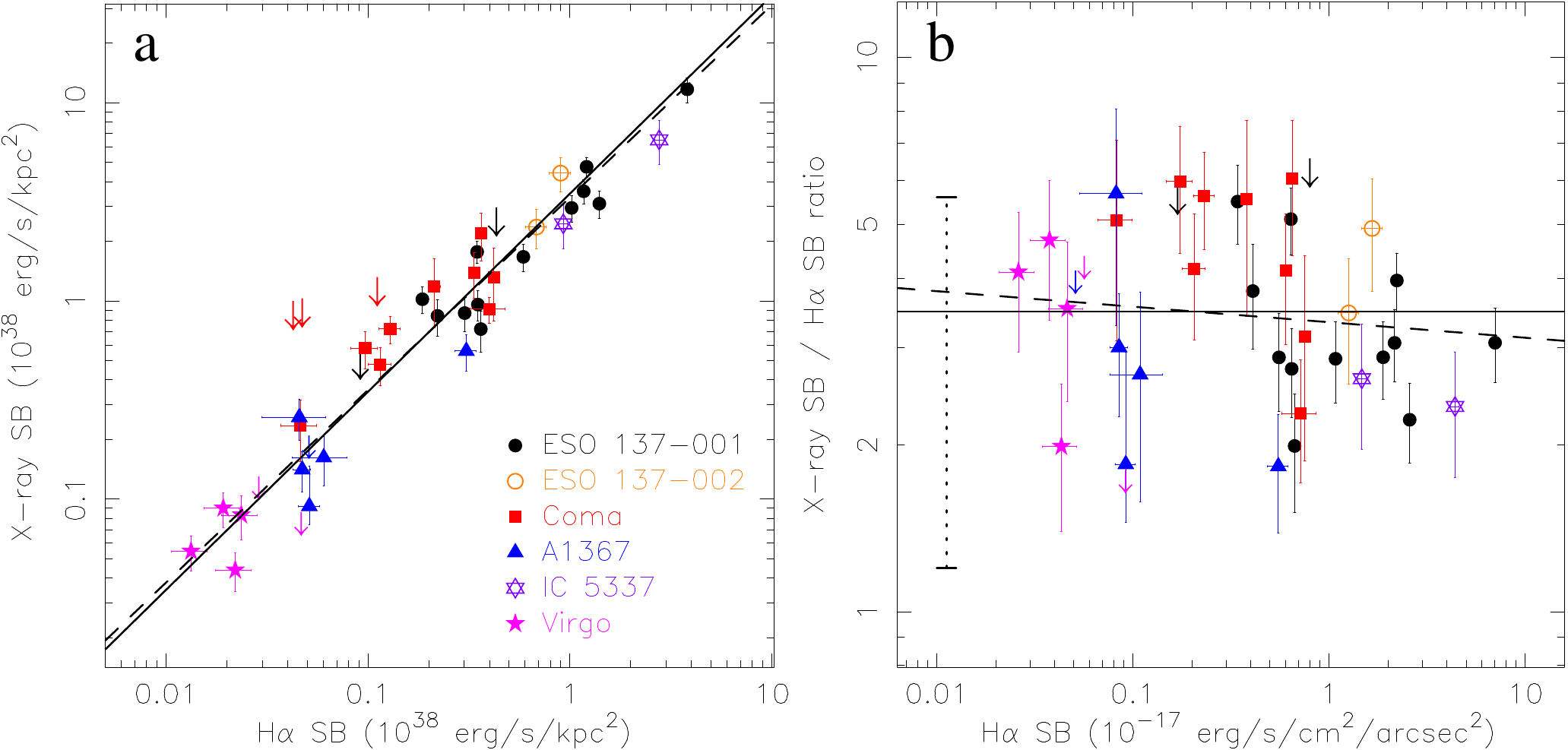}
\caption{\textbf{H$\alpha$ --- X-ray surface brightness (SB) correlation for diffuse gas in stripped tails.} 
X-ray luminosity and flux are bolometric. Errors for detections are 1-$\sigma$. 
Upper limits are 5-$\sigma$.
{\bf a}, the correlation shown as the luminosity SB.
The black dashed line shows the best fit from the Bayesian method developed by \protect\citemain{Kelly07}, with upper limits included (but excluding three upper limits in Coma as they are not constraining), SB$_{\rm X}$ = (3.33$\pm$0.34) SB$_{\rm H\alpha}^{0.97\pm0.05}$.
The black solid line shows the best fit with a slope of 1 from the same method, or SB$_{\rm X}$ / SB$_{\rm H\alpha}$ = 3.48$\pm$0.25.
{\bf b}, the correlation shown as the flux SB ratio vs. the H$\alpha$ flux SB. The solid and dashed lines are the same best fits as those in the left panel. In the ratio plot, three Coma tails with the ratio upper limits of higher than 10 are excluded as they are not constraining. The dotted line shows the ratios of 1.2 - 5.6 derived by \protect\citemain{Strickland02} for several regions in the galactic wind of NGC~253.
}
\label{fig:relation1}
\end{center}
\end{figure*}

\newpage 
\begin{figure*}[!htbp]
\begin{center}
\vspace{-0.2cm}
\includegraphics[angle=0,width=0.88\columnwidth]{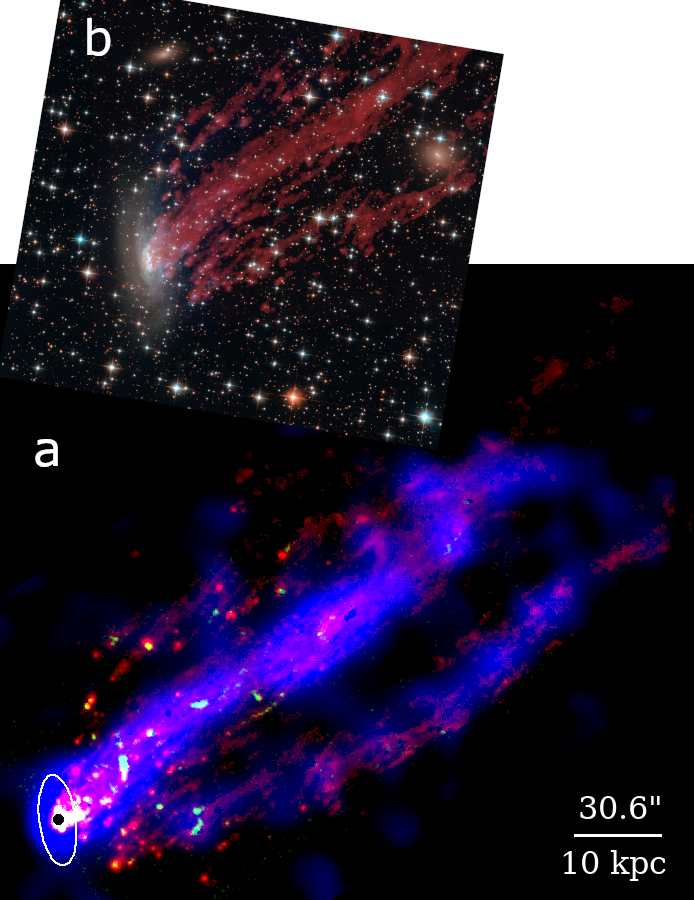}
\vspace{-0.4cm}
\caption{\textbf{The H$\alpha$ --- X-ray correlation for \ga's tail with the brightest X-ray and H$\alpha$ emission.}
{\bf a}, the X-ray emission is shown in blue (0.6 - 2 keV from \protect\citemain{Sun10}) and the H$\alpha$ emission from this work is shown in red, while the CO (2-1) emission from {\em ALMA} (\protect\citemain{Jachym19}) is shown in green. The nucleus of the galaxy is marked with a black dot and the ellipse in white shows the half-light region of the galaxy from its \hst{} F160W data.
One can also see a very good correlation between the H$\alpha$ and the X-ray diffuse emission in the tail.
{\bf b}, the {\em HST} composite image of the galaxy (Credit: STScI) is also shown with the {\em MUSE} H$\alpha$ in red.
}
\label{fig:eso137001}
\end{center}
\end{figure*}

\newpage

\begin{methods}

We assume that $\rm H_{0} = 70$ km s$^{-1}$ Mpc$^{-1}$, $\Omega_{\rm M} = 0.3$ and $\Omega_{\rm \Lambda} = 0.7$. The distance to the Virgo cluster is 16.5 Mpc \citemain{Mei07} and the distance to the other clusters is derived from the assumed cosmological parameters. At the distance of the Virgo cluster, A3627, A1367 and Coma, 1$''$ corresponds to 80 pc, 327 pc, 445 pc and 466 pc, respectively.
All X-ray spectra are fitted using the C-statistic, and the quoted uncertainties are 1$\sigma$. 

\begin{table*}
\small
\begin{center}
\caption{The sample of the RPS galaxies in this work}
\label{table:galaxy1}
\vspace{0.3cm}
\begin{center}
\begin{tabular}{lcccccc}
\hline
\hline
  {Galaxy (Cluster)} & {$z$} & {$L_{\rm W1}$ $^{a}$} & {SFR $^{b}$} & {$L_{\rm FIR}$ $^{c}$} & {$r$ $^{d}$} & {\# $^{e}$} \\ 
  & & (10$^{9}$ L$_{\odot}$) & (M$_{\odot}$/yr) & (10$^{9}$ L$_{\odot}$) & (kpc) & \\
\hline  
NGC~4569 (Virgo) & -0.00078 & 0.74 & 1.3 & 9.9$^{+0.3}_{-0.8}$ & 477 & 5+1 \\
NGC~4330 (Virgo) & 0.00521 & 0.058 & 0.17 & 1.2$\pm$0.2 & 607 & 1+1 \\
\ga\ (A3627) & 0.0154 & 0.49 & 0.97 & 5.2$\pm$0.1 & 172 & 14+2 \\
\gb\ (A3627) & 0.0190 & 3.4 & 1.1 & 12.2$\pm$0.2 & 97 & 2+3 \\
UGC 6697 (A1367) & 0.0224 & 2.4 & 6.2 & 23$\pm$1 & 614 & 2+1 \\
CGCG 097-073 (A1367) & 0.0243 & 0.18 & 1.1 & 4.4$\pm$0.3 & 865 & 1+1 \\
CGCG 097-079 (A1367) & 0.0236 & 0.41 & 1.6 & 6.1$\pm$0.2 & 816 & 1+1 \\
2MASX J11443212+2006238 (A1367) & 0.0240 & 1.0 & 3.5 & - & 703 & 1+1 \\
CGCG 097-092 (A1367) & 0.0213 & 0.68 & 1.6 & - & 886 & 1+0 \\
D100 (Coma) & 0.0171 & 0.55 & 0.50 & 4.3$\pm$0.1 & 236 & 4+1 \\
NGC~4848 (Coma) & 0.0235 & 2.7 & 4.0 & 26.1$\pm$0.7 & 764 & 3+1 \\
IC~4040 (Coma) & 0.0262 & 1.9 & 2.9 & 27.9$\pm$0.5 & 406 & 1+1 \\
GMP~2923 (Coma) & 0.0289 & 0.083 & 0.10 & - & 346 & 1+0 \\
GMP~3779 (Coma) & 0.0185 & 0.82 & 1.2 & 6.9$\pm$0.2 & 554 & 1+1 \\
GMP~3816 (Coma) & 0.0315 & 0.91 & 1.5 & 9.8$\pm$0.2 & 362 & 1+1 \\
GMP~4555 (Coma) & 0.0277 & 0.60 & 0.81 & 5.7$\pm$0.2 & 656 & 1+1 \\
IC~5337 (A2626) & 0.0619 & 10  & 7.4 &  - & 84 & 2+1 \\
\hline
\end{tabular}
\end{center}
\begin{tablenotes}
\item {\sl Note.}
$^a$: The {\em WISE} band 1 (3.6 $\mu$m) luminosity as a proxy of the stellar mass. The Galactic extinction was corrected with the relation from \citemain{2005ApJ...619..931I}. The typical errors are 2\% - 5\%.
$^b$: The star formation rate (SFR) derived from the {\em GALEX} NUV flux density and the total far-infrared (FIR) luminosity from {\em Herschel} with the relation from \citemain{2011ApJ...741..124H}. If the FIR luminosity is unavailable, the SFR is derived from the {\em GALEX} NUV flux density and the {\em WISE} 22 $\mu$m flux density with the relation from the same reference. The Kroupa IMF is assumed and the typical rms scatter of the relation is $\sim$ 23\%.
$^c$: The total FIR luminosity derived from the {\em Herschel} data.
$^d$: The projected distance of the galaxy to the cluster X-ray center.
$^e$: The number of regions in the tail + galaxy, respectively.
\end{tablenotes}
\end{center}
\end{table*}

\normalsize

\subsection*{\chandra\ and \xmm\ data processing}

We processed the \chandra\ data with the \chandra\ Interactive Analysis of Observation (CIAO; version 4.12.1) and calibration database (CALDB; version 4.9.3), following the procedures in our previous works \citemain{Sun10,Zhang13}. We particularly take advantage of two \chandra\ large programs on the Coma cluster
(13800359, 0.5 Msec, PI: Sanders; 17800479, 1 Msec, PI: Zhuravleva). 

We processed the \xmm\ data using the Extended Source Analysis Software (ESAS) integrated into the \xmm\ Science Analysis System (SAS; version 17.0.0), following the procedures in \citemain{Ge19}. 
Briefly, we reduce the raw event files from MOS and pn CCDs using tasks {\tt emchain} and {\tt epchain}, respectively. The solar soft proton flares are filtered out with {\tt mos-filter} and {\tt pn-filter} to obtain clean event files. 

For selected tail regions, the local background from regions adjacent to the tail regions is always used. The conversion from the count rate to the X-ray flux is derived from the spectral fit of the full tail region and is assumed to be the same for different tail regions. We use the AtomDB (version 3.0.9) database of atomic data and the solar abundance table ASPL from \citemain{Asplund09}. We use the XSPEC {\tt apec} model to fit the X-ray emission from tails. For the \xmm\ data, the spectra from MOS/pn are fitted jointly. The Galactic column density is modeled using {\tt tbabs}, with values from the NHtot tool \citemain{Willingale13}. X-ray flux and luminosity shown are bolometric (0.01 - 100 keV).

\subsection*{H$\alpha$ data processing}

For six galaxies in the sample (\ga, UGC~6697, CGCG~097-073, CGCG~097-079, D100 and IC~5337), there is complete or partial coverage of their tails with {\em MUSE}.
{\em MUSE} \citemain{2010SPIE.7735E..08B} is an optical integral-field spectrograph on the Very Large Telescope, which provides a $1'\times1'$ field of view with wavelength coverage from 4750 \AA\ to 9350 \AA\ in the nominal mode.
For \ga, {\em MUSE} observations come from the European Southern Observatory (ESO) programs 60.A-9349(A) (Science Verification), 60.A-9100(G) (Science Verification), 095.A-0512(A) (PI: M. Sun) and 0104.A-0226(A) (PI: M. Sun), with 12 pointings from June 21, 2014 to March 17, 2020 to cover the full tail.
The total exposure time is 5 hours and 35 minutes, with the tail regions observed in 20.9 - 35 minutes and some galaxy regions observed in 90 minutes. The seeing is $0.57''$ - $1.94''$ with a median of $1.04''$.
The {\em MUSE} data effectively remove the Galactic stars and the resulting H$\alpha$ image is much deeper than the narrow-band image by \citemain{Sun07b}. This new mosaic also covers the tail much more completely than the old one published in \citemain{Fumagalli14,Fossati16}.
D100 was observed on May 11, 2019, under the ESO program 0103.A-0684(A) (PI: P. J\'{a}chym),
with a median seeing value of $1.12''$.
Only one field was observed with a total exposure time of 33 minutes to cover the galaxy and the front part of the tail.
We also used the H$\alpha$ flux maps measured with {\em MUSE} on three A1367 galaxies, UGC~6697, CGCG~097-073 and CGCG~097-079, under the ESO programs 096.B-0019(A) and 098.B-0020(A) (PI: G. Gavazzi). The H$\alpha$ flux map of UGC~6697 comes from \citemain{Consolandi17} and detail of the observations can be found there. CGCG~097-073 was observed in three pointings, with 43 minutes on the galaxy and 38.85 minutes on two
tail regions. CGCG~097-079 was observed in five pointings, with 43 minutes on the galaxy and 38.85 minutes on four tail regions. The observations of these two galaxies were taken from Feb. 12, 2016 to Apr. 1, 2017, with seeing of $0.35''$ - $1.31''$ (a median of $0.76''$).

We used the {\em MUSE} pipeline (version 2.8.1 \citemain{2020A&A...641A..28W}) to reduce the raw data.
The {\em MUSE} pipeline was run with the ESO Recipe Execution Tool (EsoRex), 
which provides a standard procedure to reduce the individual exposures and combine them into a datacube.
The additional sky subtraction was performed with the Zurich Atmosphere Purge software (\citemain{2016MNRAS.458.3210S}).
To combine the individual datacubes
into the final mosaic, we adopted the CubeMosaic class implemented in the MUSE Python Data Analysis Framework (MPDAF)
package \citemain{2016ascl.soft11003B}. 
For the final datacube mosaic, we adopted the public IDL software Kubeviz
\citemain{Fossati16} to perform the spectral analysis. We first used the colour excess to correct the Galactic extinction, which was
obtained from the recalibration \citemain{2011ApJ...737..103S} of the dust map of \citemain{1998ApJ...500..525S} and adopted a Galaxy extinction law from
\citemain{1999PASP..111...63F} with $R_{V}=3.1$. 
We also smoothed the datacube with a Gaussian kernel of 4 pixels (or 0.8$''$) for \ga\ and 3 pixels (or 0.6$''$) for D100. We fitted the Gaussian profile to each emission line and obtained the flux. The spaxels with
S/N $<$ 3 and velocity error and velocity dispersion error $>$ 50 km s$^{-1}$ are masked. The detailed {\em MUSE} results will be presented in future papers.

Other H$\alpha$ data come from the narrow-band imaging observations as listed in Supplementary Table~\ref{table:data1}. 
\NII\ emission is removed by assuming \NII\ $\lambda$6584 / H$\alpha$ = 0.4 and \NII\ $\lambda$6548 / \NII\ $\lambda$6584 = 1/3. Galactic extinction is corrected in all cases as stated above. Bright \hii\ regions in the tail are masked for both the {\em MUSE} and narrow-band imaging data. The same criteria as used in \citemain{Fossati16} are applied for the selection of \hii\ regions. Specifically, SEXTRACTOR was run on H$\alpha$ images. \hii\ regions are selected as point-like sources (CLASS\_STAR$>$0.9) with a low ellipticity ($e<$0.2).
For diffuse H$\alpha$ emission in tails, no correction on any intrinsic extinction was applied, for both narrow-band imaging data and the {\em MUSE} data. The correction on the stellar absorption was made on the galaxy for both {\em MUSE} and narrow-band imaging data. Such a correction affects little on our results on tails since the stellar continuum beyond $D_{25}$ is very weak.

\subsection*{Data availability}
The X-ray and optical data that support the plots within this paper and other findings of this study are either publicly released (\chandra, \xmm\ and {\em VLT}/{\em MUSE} data) or published (narrow-band imaging data), as shown in the Supplementary Table~\ref{table:data1}. The key results of this work (X-ray and H$\alpha$ SB in tail regions) are also attached as an online table. Other results and reduced images of this work are available from the corresponding author M. S. upon reasonable request.

\subsection*{Code availability}
The software to reduce the X-ray and optical data in this work is publicly released. Upon request, the corresponding author M. S. will provide code (Python and Wip) used to produce the figures.

\bibliographystylemain{naturemag}
\bibliographymain{tails.bib}
\clearpage
\end{methods}

\setcounter{page}{1} %page number restarted from 1
\clearpage

\title{A universal correlation between warm and hot gas in the stripped tails of cluster galaxies}
\maketitle

\author{Ming Sun$^{1\star}$, 
Chong Ge$^{1}$, 
Rongxin Luo$^{1}$, 
Masafumi Yagi$^{2}$, \\
Pavel J\'{a}chym$^{3}$,
Alessandro Boselli$^{4}$,
Matteo Fossati$^{5,6}$,
Paul E.J. Nulsen$^{7,8}$, \\
Michitoshi Yoshida$^{9}$,
Giuseppe Gavazzi$^{5}$
}

\begin{affiliations}
 \item Department of Physics and Astronomy, University of Alabama in Huntsville, Huntsville, AL 35899, USA
 \item National Astronomical Observatory of Japan, 2-21-1, Osawa, Mitaka, Tokyo, 181-8588, Japan
 \item Astronomical Institute, Academy of Sciences of the Czech Republic, Bo\v{c}n\'{l} II 1401, 14100 Prague, Czech Republic
  \item Aix Marseille Universit\'{e}, CNRS, LAM (Laboratoire d’ Astro physique de Marseille) UMR 7326, 13388, Marseille, France
 \item Dipartimento di Fisica G. Occhialini, Universit\`{a} degli Studi di Milano Bicocca, Piazza della Scienza 3, I-20126 Milano, Italy
 \item INAF-Osservatorio Astronomico di Brera, via Brera 28, I-20121 Milano, Italy
 \item Center for Astrophysics \textbar{} Harvard \& Smithsonian, Cambridge, MA 02138, USA
 \item ICRAR, University of Western Australia, 35 Stirling Hwy, Crawley, WA 6009, Australia
 \item Subaru Telescope, National Astronomical Observatory of Japan, 650 North A'ohoku Place, Hilo, HI 96720, USA
\end{affiliations}

\section*{Supplementary Information}

\renewcommand{\figurename}{Supplementary Figure}
\setcounter{figure}{0}   

\setcounter{table}{0}
\renewcommand{\tablename}{Supplementary Table}

\subsection*{The RPS galaxy sample}

The RPS galaxies in the sample of this work are selected with these criteria:
1) $z < 0.025$ for the host cluster as more distant RPS tails are generally too faint in X-rays (the only exception is IC~5337 as discussed below). At higher $z$, it is also more difficult to separate the bright, compact \hii\ regions from the diffuse H$\alpha$ emission, and separate X-ray point sources from the diffuse X-ray emission.
2) The galaxy has a known H$\alpha$ tail beyond $D_{25}$ of the galaxy (values retrieved from HyperLeda \cite{2014A&A...570A..13M}).
3) The galaxy is covered by \chandra\ or \xmm\ and is detected in X-rays.

The initial RPS galaxy samples come from these works: \cite{Sun07bs,Sun10s,Yagi10s,Smith10s,Fossati12,Boselli16s,Yagi17s,Fossati18s}. They are in four nearby galaxy clusters, Virgo, Abell~3627, Abell~1367 and Coma, as listed in Supplementary Table~\ref{table:cluster}. Not all RPS galaxies in these samples are included because of the lack of sufficient X-ray coverage or the lack of the extended H$\alpha$ emission beyond $D_{25}$ of the galaxy. As shown in Supplementary Table~\ref{table:data1}, some Coma galaxies are covered by deep \chandra\ observations.
However, those around the cluster center (D100 and GMP~2923) are embedded in strong cluster background, which makes the detection of faint X-ray tails challenging (see discussion in ``The properties of X-ray tails'').
For the three Coma upper limits, two of them were only observed for 29.7 ks with ACIS-S in 2016 - 2017, which corresponds to only 17.2 ks with ACIS-S in cycle 5. The other upper limit is GMP~2923 that is close to the bright cluster center. Even for GMP~3816 and GMP~4555 that are covered by very deep \chandra\ ACIS-I data (977.3 ks and 846.5 ks respectively), the corresponding ACIS-S time in cycle 5 would be 413.9 ks and 307.1 ks respectively. With the low X-ray luminosities of their tails (Supplementary Table~\ref{table:galaxy2}), the detection significance is only 5.8 - 7.0 $\sigma$. 

\cite{2019ApJ...887..155Ps} also studied the H$\alpha$ - X-ray correlation in one of the GASP galaxies, IC~5337 in Abell~2626 at $z$=0.055. They presented a relation of SB$_{\rm H\alpha} \propto$ SB$_{\rm X}^{0.44\pm0.17}$ in four regions (with a limited range of surface brightness, or SB) in the stripped tail, which is different from the relation presented in this work. For comparison, we include IC~5337 in our sample with our own analysis. IC~5337 is a far more luminous galaxy than other galaxies in our sample (Table~\ref{table:galaxy1}) and it is also luminous in X-rays.
Our results in two tail regions suggest similar X-ray-to-H$\alpha$ ratios to what we have found from the local cluster sample (Fig.~\ref{fig:relation1}).
\cite{2021ApJ...911..144C} presented a similar analysis to another GASP galaxy in Abell~85 at $z$=0.055.
However, little X-ray emission is detected beyond $D_{25}$ of the galaxy so there is no H$\alpha$/X-ray correlation in the diffuse tail that can be established from the data. 

As shown in Table~\ref{table:galaxy1}, these galaxies span a wide range on their properties, factors of nearly 180 in stellar luminosity, about 40 in SFR, and over 20 in the FIR luminosity. Located in the closest cluster (Virgo) and the closest rich clusters (A3627, A1367 and Coma), the RPS galaxies in this sample are ideal for detailed analysis, compared with more distant RPS galaxies, e.g., GASP galaxies in clusters at $z$ = 0.04 - 0.07 \cite{Poggianti17s} and RPS galaxies selected from the {\em HST} observations of clusters at $z$ = 0.3 - 0.7 (e.g., \cite{Yagi15,McPartland16}). The lack of strong night sky lines between 6578\AA\ - 6827\AA\ also makes the studies of faint H$\alpha$ emission at $z <$ 0.04 easier \cite{Osterbrock96}.

As a final note, the first version of the H$\alpha$ - X-ray correlation studied in this paper was presented in the \chandra\ proposal ``Tales of tails: the Coma episode'' in March 2015 (proposal \#: 17800174, PI: Sun), with 17 regions from 4 galaxies (\ga, \gb, D100 and NGC~4848). It was also presented in an oral talk at the Symposium S8 ``Ram pressure stripping and galaxy evolution'' in the European Week of Astronomy and Space Science 2017, with the presentation url: \url{https://space.asu.cas.cz/~ewass17-soc/Presentations/S08/Sun.pdf} (page 42). While the correlation plot of SB was not shown in the published version on the above link, the X-ray/H$\alpha$ ratio mentioned in the text ($\sim$ 3.0 as a simple fit by eye) is similar to results of this paper with a larger sample and better data.

\begin{deluxetable}{lcccc}
\tablecolumns{5}
\tabletypesize{\small}
\tablecaption{The cluster sample in this work}
  \tablewidth{0pt}
  \tablehead{
  \colhead{Cluster} & \colhead{$z$} & \colhead{$kT$ $^{a}$} & \colhead{$\sigma ^{b}$} & \colhead{$r_{500}$ $^{c}$} \\
  & & (keV) & (km/s) & (kpc)
}
\startdata
Virgo & 0.00360 & 2.5 & 638 & 774 \\
A3627 & 0.01605 & 5.6 & 925 & 1199 \\
A1367 & 0.0220 & 3.6 & 726 & 938 \\
Coma & 0.0231 & 8.1 & 873 & 1465 \\
A2626 & 0.0551 & 3.2 & 650 & 866 \\
\enddata
\begin{tablenotes}
\item {\sl Note.}
$^a$: cluster temperatures from \cite{2001ApJ...549..228S}, \cite{Ikebe02} and \cite{2008A&A...478..615S};
$^b$: radial velocity dispersion of the cluster from \cite{2020A&A...635A.135K}, \cite{Woudt08}, \cite{Sohn20} and \cite{2017A&A...607A..81B};
$^c$: $r_{500}$ (the radius enclosing an overdensity of 500 times the critical density of the Universe and is $\sim$ 2/3 of the virial radius) derived from the $M - T$ relation from \cite{Sun09}.
\end{tablenotes}
\label{table:cluster}
\end{deluxetable}

\subsection*{The X-ray and H$\alpha$ data}

The X-ray and H$\alpha$ data used in this work are summarized in Supplementary Table~\ref{table:data1}.
The H$\alpha$ data come from both the IFS data from {\em VLT}/{\em MUSE} and narrow-band imaging data. Since the IFS data provide a much more robust continuum subtraction and flux calibration than what is generally achieved from the narrow-band imaging data, we also compare the results from both sets of the H$\alpha$ data. We first compare the H$\alpha$ flux for regions with both {\em MUSE} and narrow-band imaging data, for \ga, UGC~6697, CGCG~097-073, CGCG~097-079 and D100. Assuming \NII\ $\lambda$6584 / H$\alpha$ = 0.4 and \NII\ $\lambda$6548 / \NII\ $\lambda$6584 = 1/3, the H$\alpha$ fluxes from {\em MUSE} and narrow-band imaging agree at $\sim$ 32\% on average. Note that if we instead use \NII\ $\lambda$6584 / H$\alpha$ = 0.2 (e.g., for tails of CGCG~097-073 and CGCG~097-079 from our analysis) and 0.7 (e.g., for some portion of D100's tail from our analysis), the H$\alpha$ flux would increase and decrease by 21\% respectively.
For the 39 regions used in the fits in Fig.~\ref{fig:relation1}, 23 regions have the H$\alpha$ flux from {\em MUSE} and the best-fit X-ray-to-H$\alpha$ ratio to these 23 regions is 3.51$\pm$0.31. For the other 16 regions with the narrow-band imaging data, the best-fit X-ray-to-H$\alpha$ ratio is 3.48$\pm$0.48. Thus, while the narrow-band imaging data present larger uncertainty on the H$\alpha$ flux, our results from the sample study should be little affected.

\begin{deluxetable}{lccccc}
\tablecolumns{6}
\tabletypesize{\small}
\tablecaption{X-ray and H$\alpha$ data used in this work}
  \tablewidth{0pt}
  \tablehead{
  \colhead{Galaxy} & \colhead{X-ray data} & \colhead{Obs ID} & \colhead{Date} & \colhead{Total (Clean) time} & \colhead{H$\alpha$ data} \\
  & & & & (ksec) &
}
\startdata

NGC~4569 & \xmm\ & 0200650101 & 2004-12-13 & 56.4 (49.8, 17.8) & \cite{Boselli16s} \\
NGC~4330 & \xmm\ & 0651790301 & 2010-06-02 & 31.1 (17.5, 9.7) & \cite{Fossati18s} \\
\ga\ & \chandra\ & 9518 & 2008-06-13 & 141.1 (139.9) & This work \\
\gb\ & \chandra\ & 12950 & 2011-01-10 & 89.9 (89.9) & \cite{Sun10s} \\
UGC~6697 & \xmm\ & 0005210101 & 2001-11-22 & 32.9 (26.1, 14.9) & \cite{Yagi17s} \& \cite{Consolandi17s} \\
         &       & 0602200101 & 2009-05-27 & 24.6 (23.6, 14.0) &  \\
         &       & 0823200101 & 2018-06-01 & 71.6 (66.8, 50.4) & \\
CGCG 097-073 & \xmm\ & 0005210101 & 2001-11-22 & 32.9 (26.1, 14.9) & \cite{Yagi17s} \& This work \\
CGCG 097-079 & \xmm\ & 0005210101 & 2001-11-22 & 32.9 (26.1, 14.9) & \cite{Yagi17s} \& This work \\
2MASX J11443212 & \xmm\ & 0823200101 & 2018-06-01 & 71.6 (66.8, 50.4) & \cite{Gavazzi17} \\
CGCG 097-092 & \xmm\ & 0823200101 & 2018-06-01 & 71.6 (66.8, 50.4) & \cite{Yagi17s} \\
D100 & \chandra\ & 9714 & 2008-03-20 & 29.7 (29.4) & \cite{Yagi10s} \& This work \\
     &           & 13993 & 2012-03-21 & 39.6 (39.3) &  \\
     &           & 13994 & 2012-03-19 & 82.0 (81.7) & \\
     &           & 13995 & 2012-03-14 & 63.0 (62.2) & \\
     &           & 13996 & 2012-03-27 & 123.1 (122.5) & \\
     &           & 14406 & 2012-03-15 & 24.8 (24.5) & \\
     &           & 14410 & 2012-03-22 & 78.5 (78.5) & \\
     &           & 14411 & 2012-03-20 & 33.6 (33.6) & \\
     &           & 14415 & 2012-04-13 & 34.5 (34.5) & \\
NGC~4848 & \chandra\ & 8188 & 2007-03-14 & 28.7 (28.7) & \cite{Fossati12} \\
         &           & 18234 & 2017-03-29 & 19.8 (19.8) & \\
         &           & 20049 & 2017-03-29 & 19.8 (19.8) & \\
         &           & 20050 & 2017-03-29 & 13.9 (13.9) & \\
         &           & 20051 & 2017-03-31 & 14.9 (14.9) & \\
         &           & 20052 & 2017-04-16 & 23.7 (23.7) & \\
IC~4040  & \chandra\ & 18235 & 2017-03-21 & 29.7 (29.7) & \cite{Yagi10s} \\
GMP~2923 & \chandra\ & 13993 & 2012-03-21 & 39.6 (39.3) & \cite{Yagi10s} \\
         &           & 13995 & 2012-03-14 & 63.0 (62.2) & \\
         &           & 14406 & 2012-03-15 & 24.8 (24.5) & \\
         &           & 14410 & 2012-03-22 & 78.5 (78.5) & \\
GMP~3779 & \chandra\ & 18237 & 2016-11-03 & 9.9 (9.9) & \cite{Yagi10s} \\
         &           & 19911 & 2016-11-03 & 9.9 (9.9) & \\
         &           & 19912 & 2016-11-07 & 9.9 (9.9) & \\
GMP~3816 & \chandra\ & 18271 & 2017-03-15 & 54.4 (54.4) & \cite{Yagi10s} \\
        &           & 18272 & 2016-03-08 & 19.8 (19.8) & \\
        &           & 18273 & 2017-02-15 & 28.3 (28.3) & \\
        &           & 18274 & 2017-03-06 & 46.5 (46.0) & \\
        &           & 18275 & 2016-03-16 & 49.4 (49.4) & \\
        &           & 18276 & 2016-03-05 & 84.2 (83.8) & \\
        &           & 18761 & 2016-03-13 & 47.4 (46.9) & \\
        &           & 18791 & 2016-03-08 & 34.6 (34.6) & \\
        &           & 18792 & 2016-03-09 & 21.3 (21.2) & \\
        &           & 18793 & 2016-03-10 & 76.9 (76.9) & \\
        &           & 18794 & 2016-03-14 & 29.3 (29.3) & \\
        &           & 18795 & 2016-03-17 & 27.8 (27.8) & \\
        &           & 18796 & 2016-03-18 & 39.6 (39.3) & \\
        &           & 18797 & 2016-03-19 & 54.4 (54.1) & \\
        &           & 18798 & 2016-03-20 & 12.9 (12.9) & \\
        &           & 19998 & 2017-03-13 & 31.7 (31.7) & \\
        &           & 20010 & 2017-02-18 & 58.3 (58.0) & \\
        &           & 20011 & 2017-02-19 & 44.5 (44.5) & \\
        &           & 20027 & 2017-03-12 & 20.8 (20.8) & \\
        &           & 20028 & 2017-04-13 & 43.5 (43.5) & \\
        &           & 20029 & 2017-03-08 & 21.8 (21.8) & \\
        &           & 20030 & 2017-04-11 & 33.2 (32.9) & \\
        &           & 20031 & 2017-03-11 & 10.9 (10.7) & \\
        &           & 20037 & 2017-03-16 & 16.9 (16.9) & \\
        &           & 20038 & 2017-03-18 & 49.9 (49.9) & \\
        &           & 20039 & 2017-03-19 & 21.9 (21.9) & \\
GMP~4555 & \chandra\ & 18236 & 2016-11-01 & 9.9 (9.9) & \cite{Yagi10s} \\
        &           & 18271 & 2017-03-15 & 54.4 (54.4) & \\
        &           & 18272 & 2016-03-08 & 19.8 (19.8) & \\
        &           & 18273 & 2017-02-15 & 28.3 (28.3) & \\
        &           & 18274 & 2017-03-06 & 46.5 (46.0) & \\
        &           & 18275 & 2016-03-16 & 49.4 (49.4) & \\
        &           & 18276 & 2016-03-05 & 84.2 (83.8) & \\
        &           & 18761 & 2016-03-13 & 47.4 (46.9) & \\
        &           & 18791 & 2016-03-08 & 34.6 (34.6) & \\
        &           & 18792 & 2016-03-09 & 21.3 (21.2) & \\
        &           & 18793 & 2016-03-10 & 76.9 (76.9) & \\
        &           & 18794 & 2016-03-14 & 29.3 (29.3) & \\
        &           & 18795 & 2016-03-17 & 27.8 (27.8) & \\
        &           & 18796 & 2016-03-18 & 39.6 (39.3) & \\
        &           & 18797 & 2016-03-19 & 54.4 (54.1) & \\
        &           & 18798 & 2016-03-20 & 12.9 (12.9) & \\
        &           & 19909 & 2016-11-01 & 9.9 (9.9) & \\
        &           & 19910 & 2016-11-02 & 9.9 (9.9) & \\
        &           & 19998 & 2017-03-13 & 31.7 (31.7) & \\
        &           & 20010 & 2017-02-18 & 58.3 (58.0) & \\
        &           & 20011 & 2017-02-19 & 44.5 (44.5) & \\
        &           & 20027 & 2017-03-12 & 20.8 (20.8) & \\
        &           & 20028 & 2017-04-13 & 43.5 (43.5) & \\
        &           & 20029 & 2017-03-08 & 21.8 (21.8) & \\
        &           & 20030 & 2017-04-11 & 33.2 (32.9) & \\
        &           & 20031 & 2017-03-11 & 10.9 (10.7) & \\
        &           & 20037 & 2017-03-16 & 16.9 (16.9) & \\
        &           & 20038 & 2017-03-18 & 49.9 (49.9) & \\
        &           & 20039 & 2017-03-19 & 21.9 (21.9) & \\
IC~5337 & \chandra\ & 3192  & 2003-01-22 & 24.8 (23.7) & \cite{2019ApJ...887..155Ps}  \\
        &           & 16136 & 2013-10-20 & 110.9 (107.9) & \\

\enddata
\begin{tablenotes}
      \item {\sl Note.}
All \chandra\ data are taken from the ACIS instrument and all \xmm\ data are taken from the EPIC instrument. The exposure time in brackets is the clean time (MOS and pn for \xmm\ respectively). The sources of the H$\alpha$ data are also listed.
\end{tablenotes}
\label{table:data1}
\end{deluxetable}

\subsection*{X-ray spectral analysis}

For each X-ray tail, a global spectrum (including all regions used for the correlation study) is extracted and fitted with a single {\tt apec} model with temperature, abundance and normalization free at the 0.4 - 7.5 keV band. 
The best-fit model also gives the conversion factor from the count rate to flux.
For X-ray tail non-detections, the 5-$\sigma$ upper limit is derived, with the assumption of a temperature of 0.9 keV and an abundance of 0.1 solar. The upper limit is derived in the H$\alpha$ tail region. The root mean square (r.m.s., or $\sigma$) of the X-ray SB is derived from the surroundings in regions with the same size as the  H$\alpha$ tail region.
Our method to derive upper limits on the X-ray emission is also verified with two sets of data on GMP~4555. Its X-ray tail is not detected in the short 30 ks ACIS-S data but detected in the 846.5 ks ACIS-I data. Indeed, its X-ray tail is only $\sim$ 60\% brighter than the r.m.s of the short 30 ks ACIS-S data.

As stated, the adopted value of the Galactic column density ($N_{\rm H}$) is from the NHtot tool \cite{Willingale13s}, which includes the column density of the molecular gas. The $N_{\rm H}$ is then higher than the column density from the \hi\ survey. The galaxies with the highest $N_{\rm H}$ values are \ga\ and \gb, 2.83$\times10^{21}$ cm$^{-2}$ and 2.8$\times10^{21}$ cm$^{-2}$ for the adopted abundance table of ASPL. If $N_{\rm H}$ is left free in the spectral fits of the ICM spectra around \ga\ and \gb\ in A3627, the best-fit $N_{\rm H}$ values are consistent with the above values. This $N_{\rm H}$ difference from our previous works on these two galaxies (\cite{Sun06s,Sun10s,Zhang13s}) is mainly from the change in the abundance table to ASPL, which also affects the abundance model used in the Galactic absorption model.

\subsection*{The properties of X-ray tails}

We summarize the properties of X-ray tails in Supplementary Table~\ref{table:galaxy2}. The results for \ga, \gb, D100 and IC~5337 were published in \cite{Sun10s,Zhang13s}, \cite{Sanders14} and \cite{2019ApJ...887..155Ps} but here their \chandra\ data and spectra were analyzed with the most recent calibration and atomic data. 
The X-ray tail of UGC~6697 reported by \protect\cite{Sun05} with the \chandra\ data only covers the front 1/3 of the \xmm\ tail, limited by the FOV of the \chandra\ data.
Six X-ray tails are new discoveries and are first presented here. 
Temperatures of the X-ray tails are $\sim 0.8 - 1.2$ keV and the abundances from the one-$kT$ fits are generally low. While the X-ray tails of \ga\ and IC~5337 are luminous, the typical X-ray luminosity of other tails is nearly 10 times lower. \ga's H$\alpha$ image from {\em MUSE} is shown in Supplementary Fig.~\ref{fig:eso137001b} and Supplementary Fig.~\ref{fig:001_region}, along with the regions selected for correlation studies.
The X-ray and H$\alpha$ images of other tails are shown in Supplementary Fig.~\ref{fig:4569}, Supplementary Fig.~\ref{fig:002}, Supplementary Fig.~\ref{fig:6697}, Supplementary Fig.~\ref{fig:097}, Supplementary Fig.~\ref{fig:2masx}, Supplementary Fig.~\ref{fig:092}, Supplementary Fig.~\ref{fig:d100}, Supplementary Fig.~\ref{fig:4848}, Supplementary Fig.~\ref{fig:3816}, Supplementary Fig.~\ref{fig:4555} and Supplementary Fig.~\ref{fig:5337}.
The generally good spatial correlation between the X-ray and the H$\alpha$ emission is evident from these individual images. 
The stripping fronts in X-rays are also always consistent with the stripping fronts in H$\alpha$.

We can write the signal-to-noise ratio of the X-ray tail at the soft X-ray band (e.g., 0.5 - 3 keV), S/N $\propto$ $f_{\rm tail} A t$ / $\sqrt{f_{\rm total} A t}$ $\propto$ $\sqrt{f_{\rm tail} (f_{\rm tail}/f_{\rm total}) A t}$, where $f_{\rm tail}$ is the flux of the X-ray tail, $f_{\rm total}$ is the flux of the total X-ray emission (mostly from the ICM) within the aperture of the X-ray tail, $A$ is the X-ray effective area of the telescope and $t$ is the exposure time. $f_{\rm tail}/f_{\rm total}$ is essentially the fraction we show in Supplementary Table~\ref{table:galaxy2}. Thus, strong X-ray detections come from bright X-ray tails on cluster outskirts, observed with X-ray telescopes with good soft X-ray response.
The above estimate assumes a uniform local background and ignores the impact of X-ray point sources (especially for \xmm\ with limited angular resolution). The X-ray responses for X-ray tails ($kT \sim$ 1 keV) and the ICM ($kT \sim$ 2 - 8 keV in this sample) also differ. Thus, the actual S/N can be lower.

\begin{deluxetable}{lcccccc}
\tablecolumns{7}
\tabletypesize{\small}
\tablecaption{Properties of the X-ray tails}
  \tablewidth{0pt}
  \tablehead{
  \colhead{Galaxy $^{a}$} & \colhead{$l$ $^{b}$} & \colhead{$kT$ $^{c}$} & \colhead{$Z$ $^{c}$} & \colhead{$L_{\rm bol}$ $^{d}$} & \colhead{Counts/$\sigma$$^{e}$} & \colhead{Fraction$^{f}$} \\
  & (kpc) & (keV) & (solar) & (10$^{40}$ erg s$^{-1}$) &  &
}
\startdata
NGC~4569 & 80 & 0.83$^{+0.04}_{-0.05}$ & 0.37$^{+0.33}_{-0.14}$ & 7.0$\pm$0.6 & 1737/11.1 & 13\% \\
\ga\ & 80 & 0.85$^{+0.06}_{-0.04}$ & 0.07$^{+0.03}_{-0.02}$ & 27$\pm$1 & 3661/24.0 & 22\% \\
\gb\ & 40 & 1.05$^{+0.08}_{-0.05}$ & 0.41$^{+0.44}_{-0.18}$ & 2.9$\pm$0.3 & 357/9.4 & 31\% \\
UGC 6697 & 125 & 0.89$^{+0.15}_{-0.13}$ & 0.08$^{+0.06}_{-0.04}$ & 3.3$\pm$0.3 & 650/9.7 & 19\% \\
CGCG 097-079 & 100 & 0.78$^{+0.36}_{-0.20}$ & 0.09$^{+0.15}_{-0.05}$ & 3.3$\pm$0.6 & 203/5.3 & 20\% \\
2MASX J11443212 & 80 & (0.95) & (0.1) & 0.94$\pm$0.16 & 340/6.0 & 18\% \\
CGCG 097-092 & 42 & 0.98$^{+0.15}_{-0.10}$ & 0.17$^{+0.20}_{-0.10}$ & 2.2$\pm$0.3 & 251/6.3 & 19\% \\
D100 & 45 & 0.99$^{+0.09}_{-0.13}$ & 0.21$^{+0.32}_{-0.12}$ & 2.8$\pm$0.4 & 610/6.4 & 9.7\% \\
NGC~4848 & 85 & 0.76$^{+0.10}_{-0.11}$ & 0.08$^{+0.07}_{-0.04}$ & 6.2$\pm$0.8 & 333/7.7 & 24\% \\
GMP~3816 & 17 & 1.17$^{+0.19}_{-0.28}$ & 0.12$^{+0.20}_{-0.10}$ & 1.2$\pm$0.2 & 341/7.0 & 18\% \\
GMP~4555 & 24 & 0.95$^{+0.20}_{-0.14}$ & 0.14$^{+0.23}_{-0.09}$ & 1.5$\pm$0.3 & 252/5.8 & 20\% \\
IC~5337 & 48 & 1.20$^{+0.09}_{-0.11}$ & 0.19$^{+0.11}_{-0.07}$ & 59$\pm$5 & 999/10.8 & 31\% \\

\enddata
\begin{tablenotes}
\item {\sl Note.}
$^a$: Only galaxies with an X-ray tail detected are included. No X-ray tails are detected from the current data of IC~4040, GMP~2923, GMP~3779, CGCG~097-073 and NGC~4330 so they are not included here. X-ray upper limits are present for them in Fig.~\ref{fig:relation1}.
$^b$: The projected length of the X-ray tail, measured from the nucleus, is determined from the current X-ray data with varying depth.
$^c$: The best-fit temperature and abundance of the X-ray tail from a single {\tt apec} fit. 2MASX~J11443212+2006238's X-ray tail is detected but too faint for a spectral analysis so a typical spectral model for its X-ray tail is assumed. The typically low abundance likely suggests the presence of multi-$T$ components in the tail, naturally from mixing.
$^d$: The X-ray bolometric luminosity of the tail.
$^e$: The 0.5 - 3 keV net counts of the X-ray tail and its significance in $\sigma$. 
$^f$: The fraction of the X-ray emission from the tail to the total X-ray emission at the 0.5 - 3 keV band. Most X-ray emission is from the ICM and there is a small contribution of the X-ray background in this soft band.
\end{tablenotes}
\label{table:galaxy2}
\end{deluxetable}

\begin{figure*}[!htbp]
\begin{center}
\vspace{-0.1cm}
\includegraphics[angle=0,width=1.0\columnwidth]{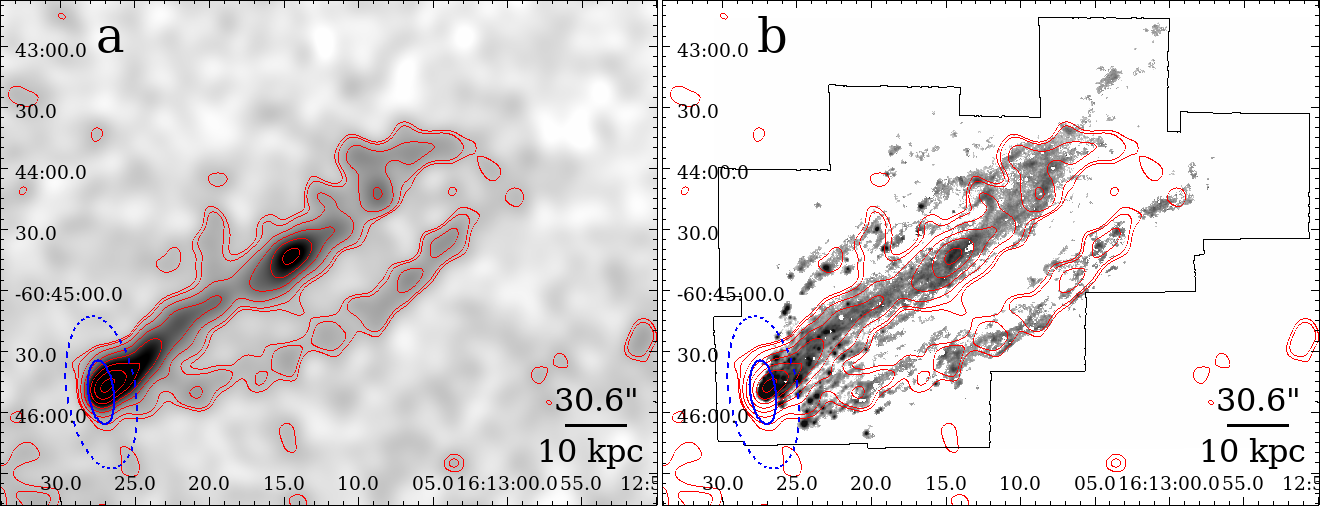}
\vspace{-0.8cm}
\caption{\textbf{X-ray and H$\alpha$ images of \ga{}.}
(a) the X-ray 0.6 - 2 keV image, (b) the net H$\alpha$ image from {\em MUSE}.
Contours of the X-ray emission (0.76, 0.80, 0.89, 1.05, 1.27, 1.55, 1.89 and 2.30 $\times10^{-5}$ counts sec$^{-1}$ arcsec$^{-2}$, starting from 3-$\sigma$ in square-root spacing) are also shown in red. The ellipse in blue solid line shows the half-light region of the galaxy from its \hst{} F160W data and the ellipse in blue dashed line shows the $D_{25}$ of the galaxy.
The FOV of the new {\em MUSE} mosaic is also shown on the right panel (black solid line).
Note that some voids in the H$\alpha$ image are caused by bright Galactic stars as \ga\ has a Galactic latitude of -7 deg. X-ray point sources are removed to better show the diffuse emission. 
Note that the X-ray image is heavily smoothed so the X-ray emission appears more extended than the H$\alpha$ emission. In fact, the X-ray leading edge is at the same position as the H$\alpha$ leading edge as shown in \protect\cite{Sun10s}.
While the H$\alpha$ tail extends beyond the X-ray tail shown, the analysis by \protect\cite{Sun10s} suggests faint X-ray extension beyond it.
X-ray point sources, \hii\ regions and background sources are excluded in deriving the X-ray and H$\alpha$ SB.
RA and DEC are in J2000 (same for all Supplementary Figures).
}
\label{fig:eso137001b}
\vspace{-0.1cm}
\end{center}
\end{figure*}

\begin{figure*}
\begin{center}
\includegraphics[angle=0,width=0.95\columnwidth]{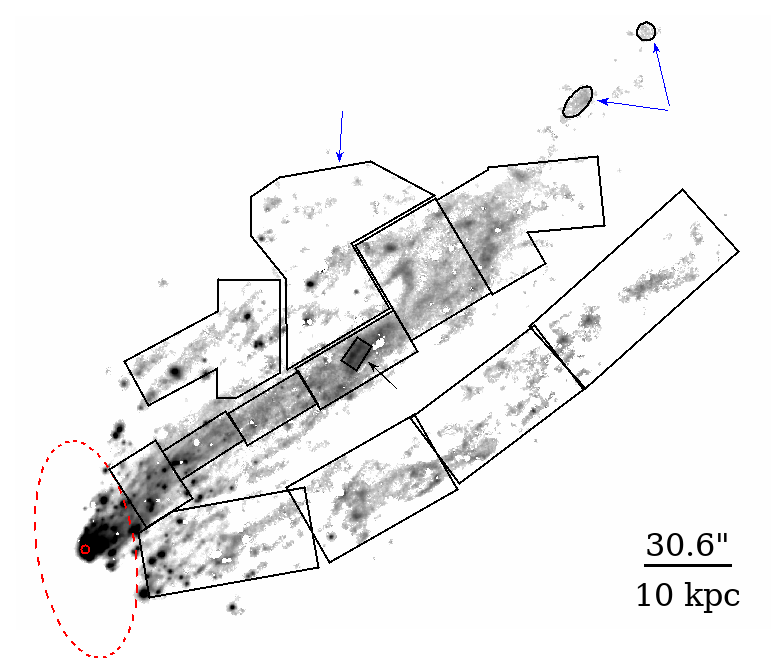}
\caption{\textbf{The selected regions on \ga's tail plotted on its H$\alpha$ image from {\em MUSE}.} The $D_{25}$ aperture of the galaxy is shown as the red ellipse (dashed line).
The region with the highest H$\alpha$ and X-ray SB in Fig.~\ref{fig:relation1} (or region \#1) is in the middle of the primary tail, marked by the black arrow. SB is also measured in the bigger region around region \#1, with region \#1 excluded. The two small ellipses beyond the primary tail, marked by blue arrows, are combined as one region with an X-ray upper limit measured. The other region with an X-ray upper limit derived is also marked by the blue arrow. For simplicity, masks for \hii\ regions, foreground stars and background sources are not shown. The nucleus of the galaxy is marked with a small red circle.
}
\label{fig:001_region}
\end{center}
\end{figure*}

\begin{figure*}
\begin{center}
\includegraphics[angle=0,width=1.0\columnwidth]{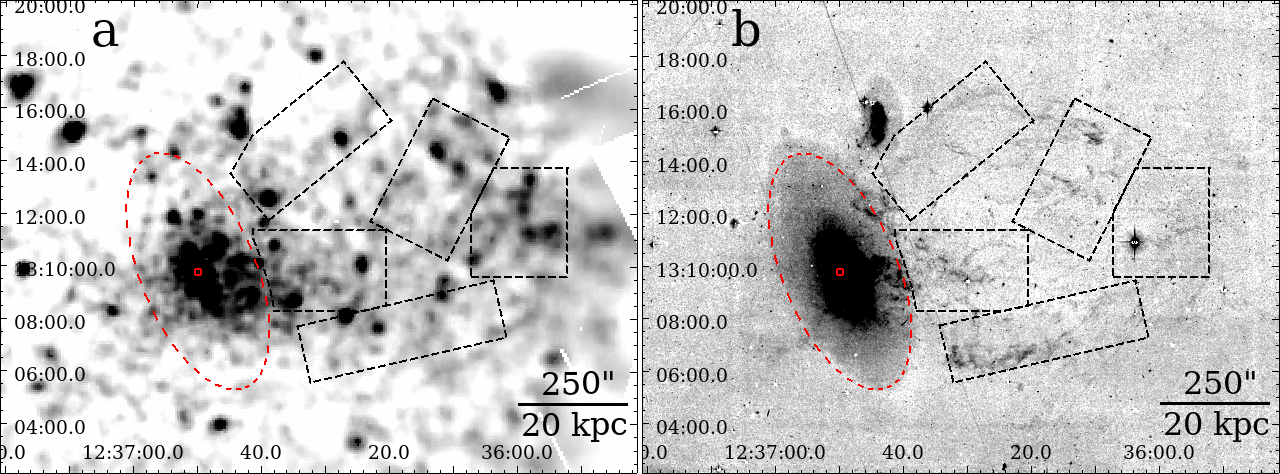}
\vspace{-0.8cm}
\caption{\textbf{X-ray and H$\alpha$ images of NGC~4569.}
(a) the X-ray 0.4 - 1.3 keV image, (b) the net H$\alpha$ image.
The $D_{25}$ aperture of the galaxy is shown as the red ellipse (dashed line), also the same for the following figures.
The nucleus position is marked by the small red circle, also the same for the following figures. Five regions (dashed line) show the tail regions (four detections and one upper limit in X-rays) used in the correlation study. 
X-ray point sources are still included in the shown image (same for the following figures unless mentioned).
}
\label{fig:4569}
\end{center}
\end{figure*}

\begin{figure*}
\begin{center}
\includegraphics[angle=0,width=1.0\columnwidth]{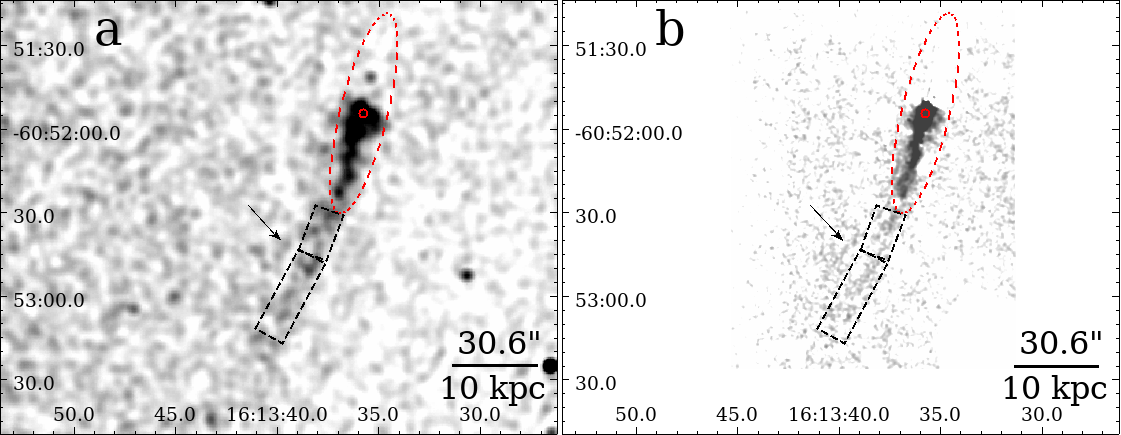}
\vspace{-0.8cm}
\caption{\textbf{X-ray and H$\alpha$ images of \gb.}
(a) the X-ray 0.6 - 2 keV image, (b) the net H$\alpha$ image. 
The two boxes (dashed line) show the tail regions used in the correlation study. The detail of these images and tails was discussed in \protect\cite{Zhang13s}. The arrow shows the location of the secondary tail. 
}
\label{fig:002}
\end{center}
\end{figure*}

\begin{figure*}
\begin{center}
\includegraphics[angle=0,width=1.0\columnwidth]{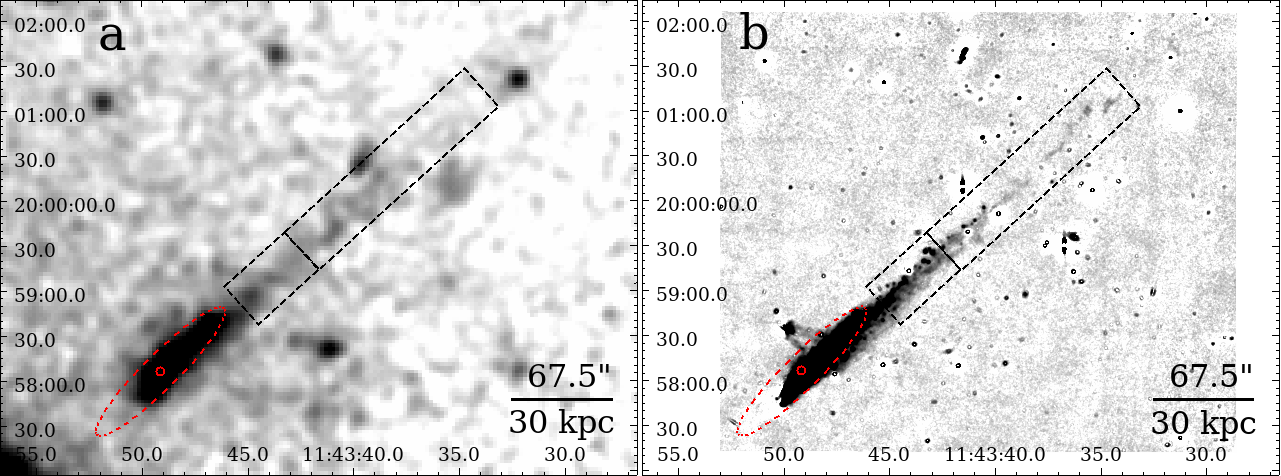}
\vspace{-0.8cm}
\caption{\textbf{X-ray and H$\alpha$ images of UGC~6697.}
(a) the X-ray 0.5 - 2 keV image, (b) the net H$\alpha$ image.
The two boxes (dashed line) show the tail regions used in the correlation study. 
}
\label{fig:6697}
\end{center}
\end{figure*}

\begin{figure*}
\begin{center}
\includegraphics[angle=0,width=1.0\columnwidth]{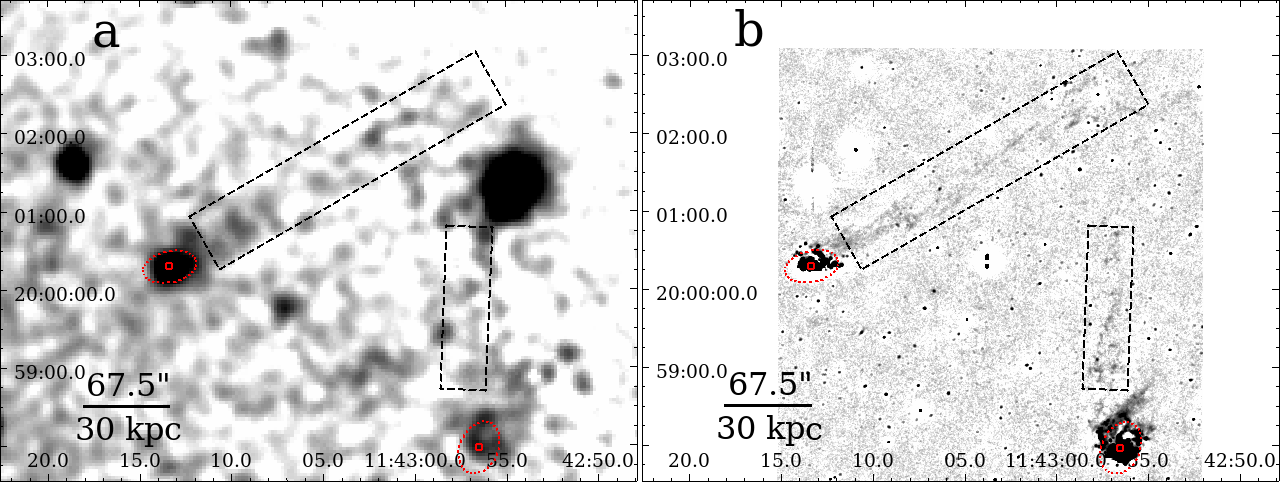}
\vspace{-0.8cm}
\caption{\textbf{X-ray and H$\alpha$ images of CGCG~097-079 (east) and CGCG~097-073 (west).}
(a) the X-ray 0.5 - 2 keV image, (b) the net H$\alpha$ image.
The boxes (dashed line) show the tail regions used in the correlation study. 
CGCG~097-073's X-ray tail is not significantly detected with the current data.
}
\label{fig:097}
\end{center}
\end{figure*}

\begin{figure*}
\begin{center}
\includegraphics[angle=0,width=1.0\columnwidth]{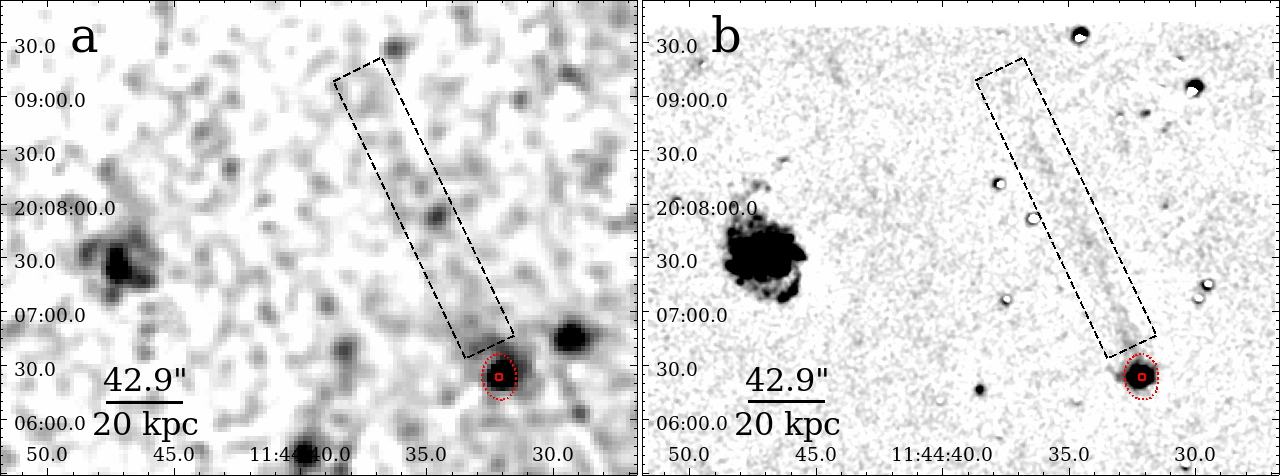}
\vspace{-0.8cm}
\caption{\textbf{X-ray and H$\alpha$ images of 2MASX~J11443212+2006238.}
(a) the X-ray 0.5 - 2 keV image, (b) the net H$\alpha$ image.
The box (dashed line) shows the tail region used in the correlation study. 
}
\label{fig:2masx}
\end{center}
\end{figure*}

\begin{figure*}
\begin{center}
\includegraphics[angle=0,width=1.0\columnwidth]{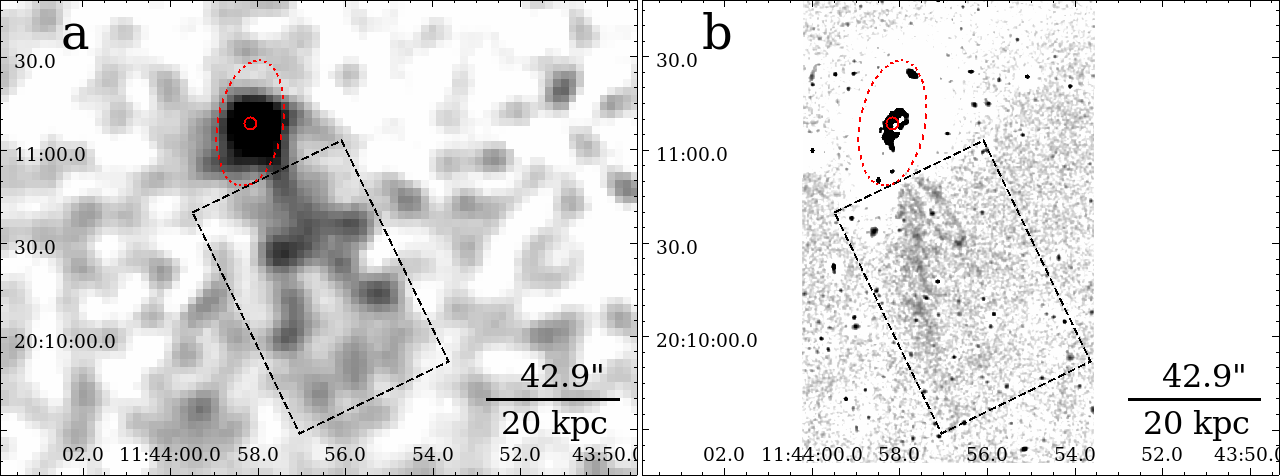}
\vspace{-0.8cm}
\caption{\textbf{X-ray and H$\alpha$ images of CGCG~097-092.}
(a) the X-ray 0.5 - 2 keV image, (b) the net H$\alpha$ image.
The box (dashed line) shows the tail region used in the correlation study. 
}
\label{fig:092}
\end{center}
\end{figure*}

\begin{figure*}
\begin{center}
\includegraphics[angle=0,width=1.0\columnwidth]{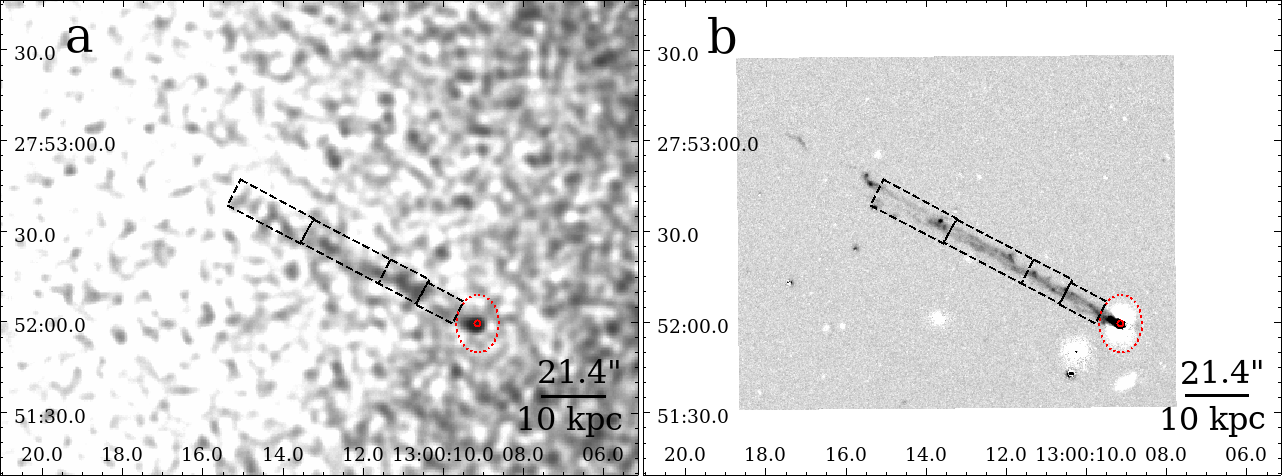}
\vspace{-0.8cm}
\caption{\textbf{X-ray and H$\alpha$ images of D100.}
(a) the X-ray 0.5 - 2 keV image, (b) the net H$\alpha$ image.
The four boxes (dashed line) show the tail regions used in the correlation study. 
}
\label{fig:d100}
\end{center}
\end{figure*}

\begin{figure*}
\begin{center}
\includegraphics[angle=0,width=1.0\columnwidth]{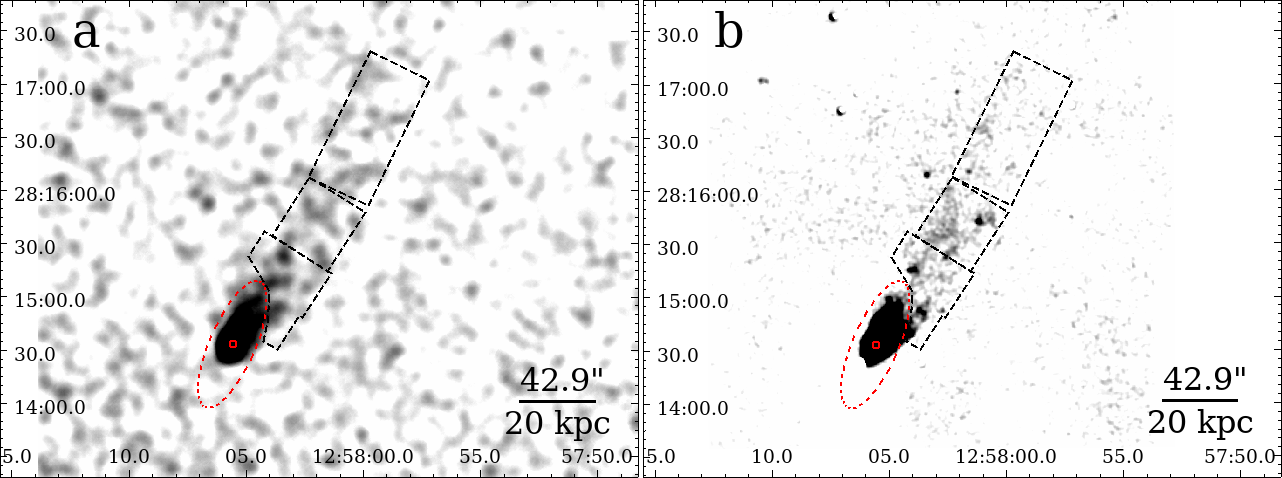}
\vspace{-0.8cm}
\caption{\textbf{X-ray and H$\alpha$ images of NGC~4848.}
(a) the X-ray 0.5 - 2 keV image, (b) the net H$\alpha$ image.
The $D_{25}$ aperture of the galaxy is shown as the red ellipse (dashed line). 
Point sources in the X-ray image is removed to show the diffuse emission better. 
}
\label{fig:4848}
\end{center}
\end{figure*}

\begin{figure*}
\begin{center}
\includegraphics[angle=0,width=1.0\columnwidth]{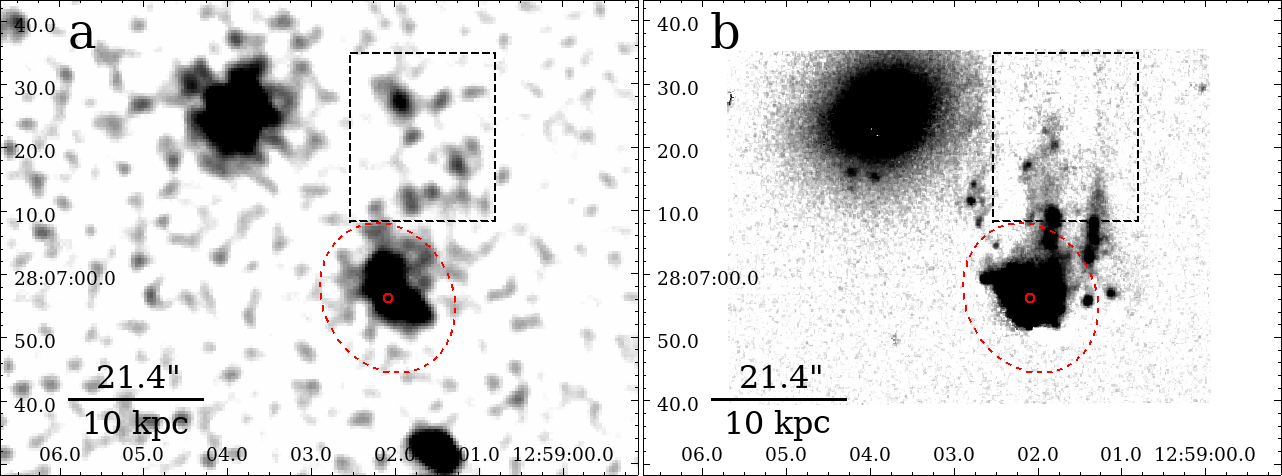}
\vspace{-0.8cm}
\caption{\textbf{X-ray and H$\alpha$ images of GMP~3816.}
(a) the X-ray 0.5 - 2 keV image, (b) the net H$\alpha$ image.
The box (dashed line) shows the tail region used in the correlation study. 
}
\label{fig:3816}
\end{center}
\end{figure*}

\begin{figure*}
\begin{center}
\includegraphics[angle=0,width=1.0\columnwidth]{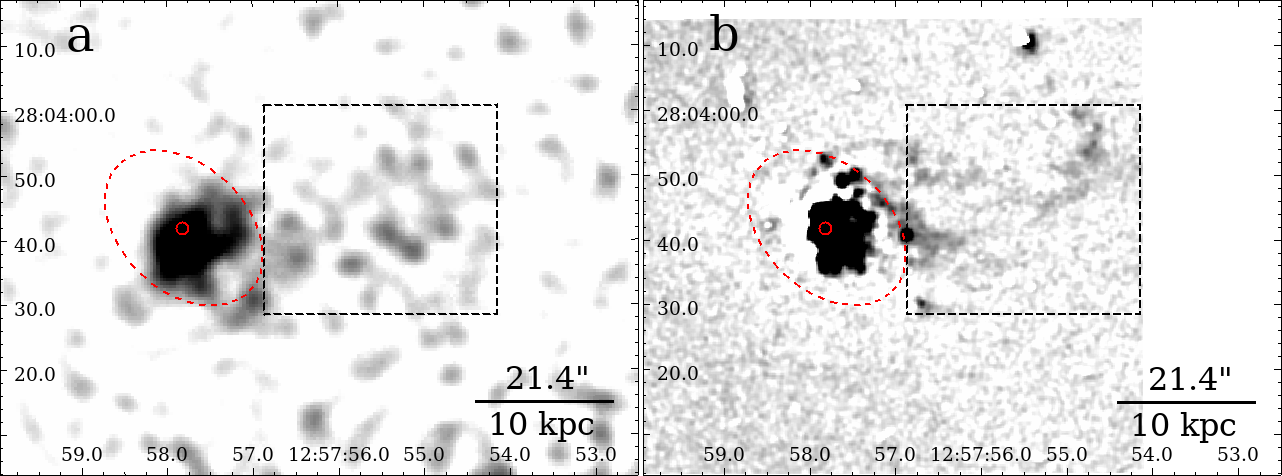}
\vspace{-0.8cm}
\caption{\textbf{X-ray and H$\alpha$ images of GMP~4555 (NGC~4858).}
(a) the X-ray 0.5 - 2 keV image, (b) the net H$\alpha$ image.
The box (dashed line) shows the tail region used in the correlation study. 
}
\label{fig:4555}
\end{center}
\end{figure*}

\begin{figure*}
\begin{center}
\includegraphics[angle=0,width=1.0\columnwidth]{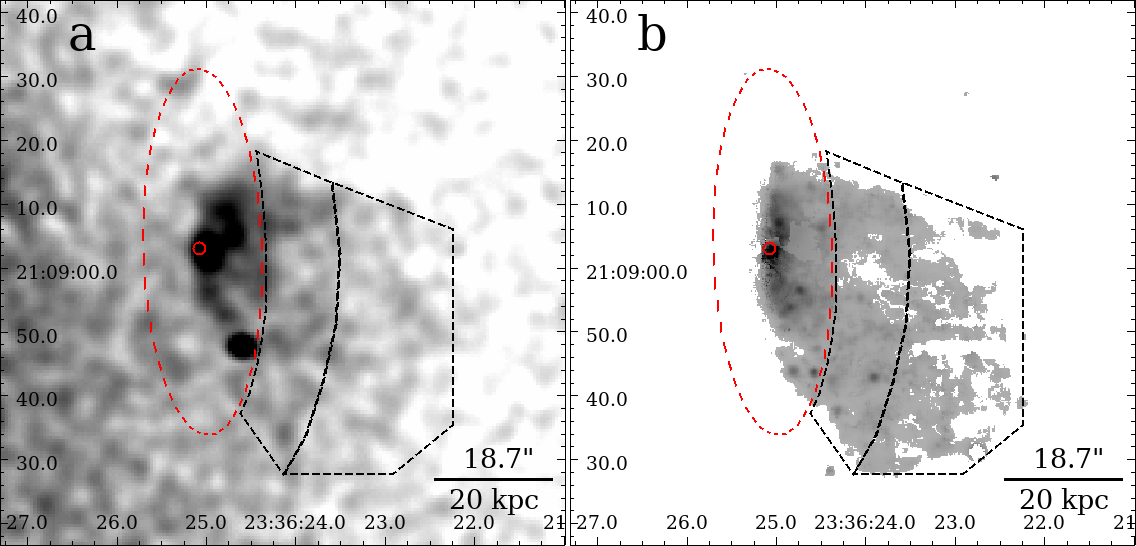}
\vspace{-0.8cm}
\caption{\textbf{X-ray and H$\alpha$ images of IC~5337.}
(a) the X-ray 0.5 - 2 keV image, (b) the net H$\alpha$ image.
The two boxes (dashed line) show the tail regions used in the correlation study. 
}
\label{fig:5337}
\end{center}
\end{figure*}

\subsection*{The effect of the region size on the relation}

Our selection of regions for the correlation study is mainly limited by the X-ray data. For many tails, only one region for the full tail can be studied. For brighter X-ray tails, regions are divided to have at least 4-$\sigma$ X-ray detection in each region. 
As the H$\alpha$ data generally have higher quality than the X-ray data, the selection is also guided by the H$\alpha$ data and we include X-ray upper limits on eight regions defined by the H$\alpha$ emission. On the other hand, there is no region in the tails with significant X-ray emission but without H$\alpha$ detection.

While it is desirable to study the X-ray --- H$\alpha$ correlation at the same physical scale for all tails, several factors make it difficult.
First, the X-ray data, as the main limiting factor, have different depths. This is determined by the combination of the telescope (\chandra\ or \xmm), exposure (for \chandra, also the time of observations as its soft X-ray response has degraded fast in recent years), Galactic absorption and the local ICM background. The last factor is important as an X-ray tail that would not be detected in a bright cluster center can easily be detected in the faint cluster outskirt with a shorter exposure. Moreover, \xmm\ has a much poorer angular resolution than \chandra\ so the adopted \xmm\ regions are large, which results in large physical sizes for regions in A1367's tails.
Second, the H$\alpha$ data also have different depths. The H$\alpha$ data for \gb\ are much shallower than those for other galaxies.
Third, even if we want to adopt the same large region size for all tails, this size cannot exceed the size of the whole tail for individual galaxies.

In Supplementary Fig.\ref{fig:relation1large}, we show the X-ray --- H$\alpha$ correlation in 31 tail regions with similar physical sizes of 300 - 440 kpc$^{2}$ (median: 352 kpc$^{2}$), with 30 detections and 1 upper limit. The twelve regions in \ga's tail with X-ray detections are grouped into four large regions. The three regions in \gd's tail are split into four regions. The two regions in UGC~6697's tail are split into four regions. The long tail of CGCG~097-079 is split into five regions. The tail of 2MASX~J11443212+2006238 is split into three regions. The tail of CGCG~097-092 is split into two regions. The two regions in IC~5337's tail are split into four regions. The five regions in \gv's tail already have area within this range. Tails of \gb, D100, GMP~3816, GMP~4555 and NGC~4330 are not included as the total area of their tails (51 kpc$^{2}$, 162 kpc$^{2}$, 130 kpc$^{2}$, 248 kpc$^{2}$ and 65 kpc$^{2}$, respectively), is less than 300 kpc$^{2}$. The three upper limits in Coma and one upper limit in A1367 are not included as they are not constraining in regions with similar physical sizes.
The ranges of the H$\alpha$ SB at the fixed 5 kpc$^2$ area in tails are also plotted near the top, which shows the different depth of the H$\alpha$ data.
As shown, the correlation still holds well in these regions with more similar sizes than those in Fig.~\ref{fig:relation1}. The best-fit relation is SB$_{\rm X}$ = (3.05$\pm$0.25) SB$_{\rm H\alpha}^{0.94\pm0.06}$. If the slope is fixed at 1, SB$_{\rm X}$ / SB$_{\rm H\alpha}$ = 3.47$\pm$0.30.
Both are consistent with results from the original set of regions used for Fig.~\ref{fig:relation1}.
Moreover, as emphasized before, the X-ray / H$\alpha$ ratios at different physical scales in tails are consistent with a constant with a small scatter, which is a robust conclusion not affected by the choice of region size.

\begin{figure*}[t]
\begin{center}
\includegraphics[angle=0,width=1.0\columnwidth]{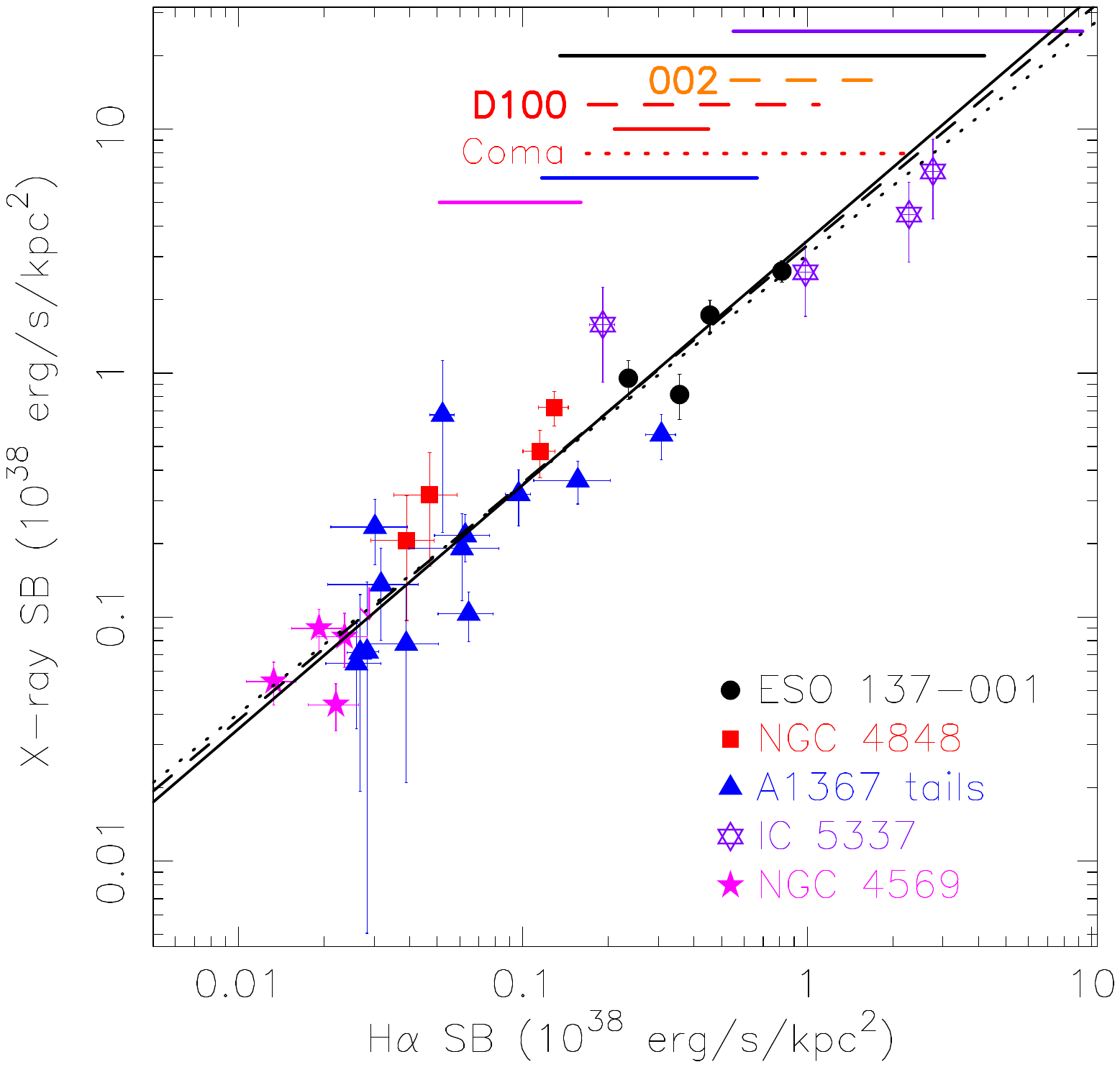}
\vspace{-1.2cm}
\caption{\textbf{H$\alpha$ --- X-ray SB correlation for diffuse gas in stripped tails, in regions with similar sizes of 300 - 440 kpc$^{2}$.} 
Errors for detections are 1-$\sigma$. Upper limits are 5-$\sigma$.
The black solid and dashed lines are still the same as those in Fig.\ref{fig:relation1}, while the dotted line shows the best fit from these regions with similar sizes.
The ranges of the H$\alpha$ SB at the fixed 5 kpc$^2$ area are also shown by straight lines near the top (orange dashed line: \gb; red dashed line: D100; red dotted line: Coma tails as in Fig.\ref{fig:relation1}).
}
\label{fig:relation1large}
\end{center}
\end{figure*}

\begin{figure*}[t]
\begin{center}
\includegraphics[angle=0,width=0.8\columnwidth]{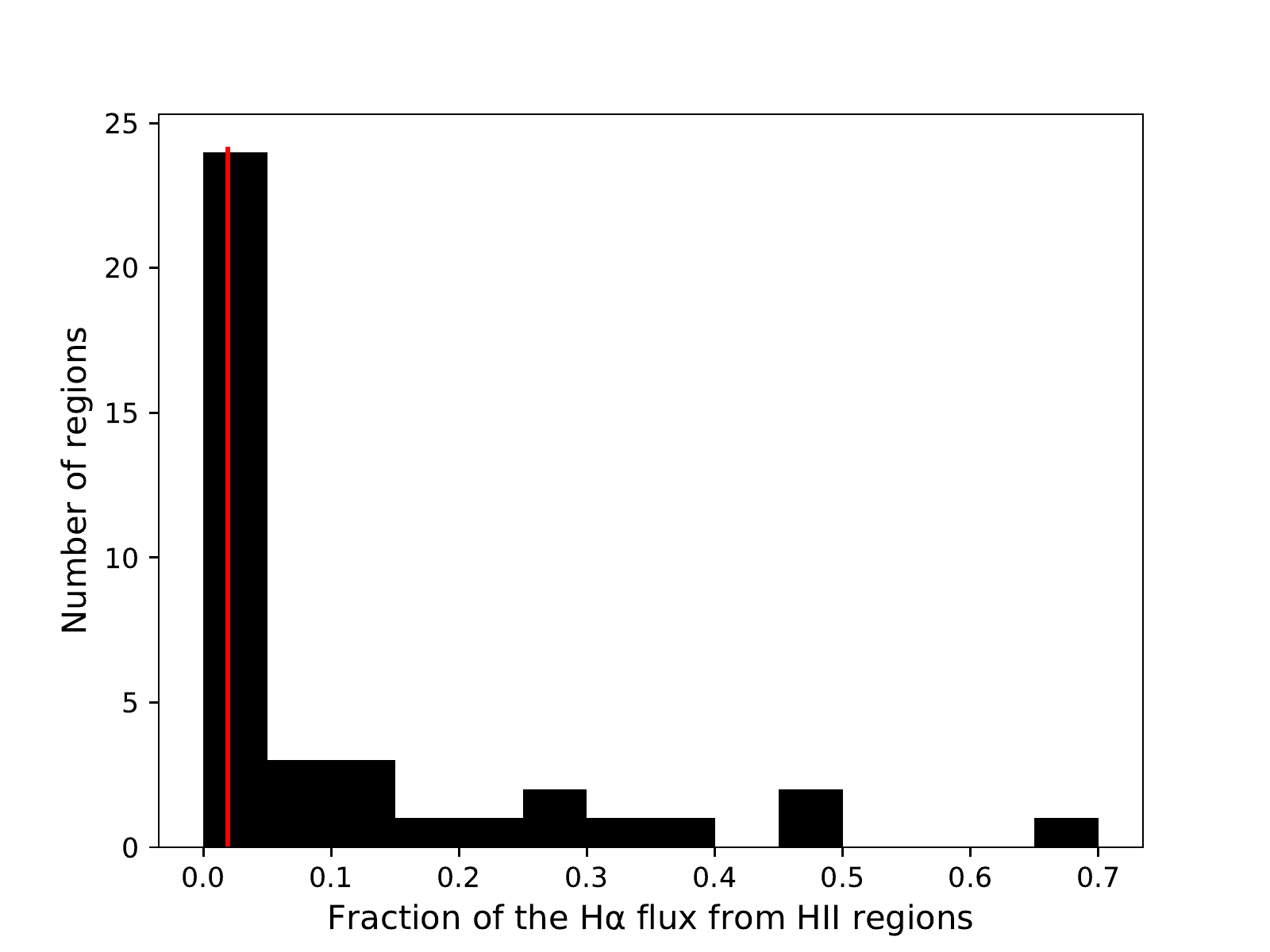}
\vspace{-0.6cm}
\caption{\textbf{The fraction of the flux from \hii\ regions to the total H$\alpha$ flux in each region used for the correlation study.} The red line shows the median value of 1.9\%.
}
\label{fig:HIIfraction}
\end{center}
\end{figure*}

\subsection*{The effect of \hii{} regions and X-ray point sources on the correlation of diffuse gas}

Our analysis focuses on diffuse gas in stripped tails so \hii{} regions are excluded. The flux fraction of \hii{} regions to the total H$\alpha$ flux in each region is calculated for each of the 39 regions used for the correlation study. The histogram is shown in Supplementary Fig.~\ref{fig:HIIfraction}. Most galaxies in our sample have weak star formation (SF) in their stripped tails. Twelve galaxies have no \hii{} regions or weak H$\alpha$ contribution ($<$ 12\% of the total flux in the region) from \hii{} regions in any region of their tails: ESO~137-002, 2MASX~J11443212+2006238, NGC~4569, D100, NGC~4848, GMP~4555, CGCG~097-092, NGC~4330, CGCG097-079, CGCG~097-073, GMP~2923 and GMP~3779. Strong SF only happens in the front part of stripped tails in \ga, UGC~6697, GMP~3816, IC~4040 and IC~5337. Even for these five galaxies, contribution from \hii{} regions to the total H$\alpha$ emission is small at regions away from the galaxy. For example, 9 regions in \ga\ have \hii{} flux fraction of less than 3.2\%.

We also examined the H$\alpha$ - X-ray correlation for regions with the \hii{} flux fraction of less than 12\%. Ten regions are then excluded so the sample includes 29 regions (25 detections + 4 upper limits). Again, the three upper limits in Coma are excluded. The fitting results are similar to the ones with 39 regions, SB$_{\rm X}$ = (3.77$\pm$0.48) SB$_{\rm H\alpha}^{1.02\pm0.06}$. If the slope is fixed at 1, SB$_{\rm X}$ / SB$_{\rm H\alpha}$ = 3.66$\pm$0.30.
If we instead include all regions with the \hii{} flux fraction of less than 1/3, we have 36 regions (31 detections + 5 upper limits). The fitting results are also similar, SB$_{\rm X}$ = (3.44$\pm$0.39) SB$_{\rm H\alpha}^{0.98\pm0.05}$. If the slope is fixed at 1, SB$_{\rm X}$ / SB$_{\rm H\alpha}$ = 3.53$\pm$0.26.
While this comparison may also suggest a small residual contamination of the \hii{} regions (or SF) in the H$\alpha$ flux of some regions in the sample, the effect is small and we can conclude that our best-fit correlation result is little affected by the uncertainty related to the removal of \hii{} regions.

While the \hii{} regions contribute significantly to the H$\alpha$ emission, their contribution to the X-ray emission is very weak. With the SFR relations from \cite{2009MNRAS.399..487R}, the soft X-ray emission from the \hii{} regions is $L_{\rm 0.5 - 2 keV} = 0.036 L_{\rm H\alpha}$. With a bolometric correction factor of $\sim$ 2 typical for the thermal X-ray emission from \hii{} regions (e.g., \cite{2009MNRAS.399..487R}), $L_{\rm X, bol} = 0.072 L_{\rm H\alpha}$. The correlation for diffuse gas in stripped tails is $L_{\rm X, bol} / L_{\rm H\alpha} \sim 3.5$ from this work, which suggests that the contribution of \hii{} regions on the X-ray emission is only $\sim$ 2\%, even if the region is full of \hii\ regions. We also ran another test to examine the impact of including the positions of \hii{} regions in the X-ray SB.
This test can only be done with the \chandra\ data for its superior angular resolution. We examined 11 regions in \ga, D100, NGC~4848, GMP~4555 and IC~5337 where the \hii{} regions contribute to more than 5\% of the total H$\alpha$ flux. X-ray SB of each region is measured with or without \hii{} regions removed. The X-ray flux ratios after and before removing \hii{} regions are 0.90 to 1.03, with a weighted average of 0.96$\pm$0.06. This again shows the very weak contribution from \hii{} regions on the soft X-ray emission. In fact, since \hii{} regions are generally close to the galaxy, an X-ray SB gradient with the distance from the galaxy can also partially account for the above ratio of less than 1.
Since the contribution of the \hii{} regions to the soft X-ray emission is very small and we cannot exclude them for the \xmm\ data without losing a large fraction of the data (large PSF of \xmm), we did not exclude the positions of the \hii{} regions in the X-ray analysis.

As stated, X-ray point sources are excluded. In fact, 14 of 17 galaxies have no X-ray point sources detected in their stripped tails. The X-ray point sources in NGC~4848 and NGC~4569 are most likely background sources and they are far away from \hii{} regions (no \hii{} regions in NGC~4569's tail anyway). Only \ga\ has $\sim$ 5 X-ray point sources close to \hii{} regions but there is always an offset between \hii\ regions and the X-ray point sources \cite{Sun10s}. Since the detected X-ray point sources are excluded and local background from the immediate surroundings is always used, the unresolved X-ray point source population should be well subtracted on average. On the other hand, tails with strong SF may have enhanced activity of high-mass X-ray binaries and even ultra-luminous X-ray sources (e.g., \cite{Sun10s}). Again with the relations established by \cite{2009MNRAS.399..487R} and assuming a power law index of 1.7, the contribution from X-ray point sources at 0.5 - 2 keV is $L_{\rm 0.5 - 2 keV} = 0.022 L_{\rm H\alpha}$, which is very small. We further performed a test by adding a power law component on the spectral fits for nine stripped tails with the best data in our sample. On average, this reduces the X-ray flux of the thermal gas by $\sim$ 11\%. However, one should be aware that this reflects more on the multi-temperature distribution in the stripped tails, rather on the embedded population of X-ray binaries (e.g., \cite{Sun10s}). Thus, we conclude that our best-fit correlation result is also robust to the uncertainty related to the removal of X-ray point source population.

\subsection*{The linear regression fits of the H$\alpha$ - X-ray correlation}

We emphasize that the correlation is not the consequence of a distance - distance correlation. In this work, we define the SB, for both H$\alpha$ and X-ray as shown in the left panel of Fig.~\ref{fig:relation1}, as the luminosity SB in units of erg s$^{-1}$ kpc$^{-2}$, or the total luminosity divided by the physical area of the region.
It is proportional to the flux SB per solid angle, in units of erg s$^{-1}$ cm$^{-2}$ arcsec$^{-2}$, for these nearby systems where the difference between the luminosity distance and the angular diameter distance is small.
Indeed, as shown in the right panel of Fig.~\ref{fig:relation1}, the flux ratios of the X-ray and H$\alpha$ SB are consistent with a constant.
While the fits based on the luminosity SB can in principle be different from fits based on flux SB, the difference for our sample of nearby galaxies is always very small so all fits presented in this paper are based on luminosity SB.

For the linear regression fit of the H$\alpha$ - X-ray correlation shown in Fig.~\ref{fig:relation1}, we used the Bayesian method developed by \cite{Kelly07s} (linmix) that can also handle the censored data (or limits).
We used its python version developed by Josh Meyers.
If only detections are fit, the best-fit relation, SB$_{\rm X}$ = (3.20$\pm$0.30) SB$_{\rm H\alpha}^{0.96\pm0.05}$, is basically the same as the one with upper limits also fit. We also attempted the BCES method by \cite{AB96} (only for detections) and derived:
SB$_{\rm X}$ = (3.30$\pm$0.25) SB$_{\rm H\alpha}^{0.97\pm0.04}$ for both the Orthogonal estimator and the Bisector estimator.
All these fits from different methods are consistent with each other.
If upper limits are included (but excluding three Coma upper limits), the intrinsic scatter derived from linmix is 9\%$\pm$3\%.
If only detections are included in the fit, the intrinsic scatter derived from linmix is 7\%$\pm$2\%.
If upper limits are included (but excluding three Coma upper limits) and the slope is forced at 1, the intrinsic scatter is 10\%$\pm$4\%.
While linmix does not provide a measure of the goodness of the fit, as shown in Fig.~\ref{fig:relation1}, the fit with a constant X-ray/H$\alpha$ ratio is essentially as good as the best-fit discussed above, when the scatter is considered.

\subsection*{The H$\alpha$ - X-ray correlation of the galaxy regions}

We also examined the H$\alpha$ - X-ray correlation in the galaxy, or more precisely within $D_{25}$ of the galaxy. There are two galaxy regions for \ga\ and three galaxy regions for \gb. Each of the other galaxies only has one galaxy region. For regions in the galaxy, no mask is applied for \hii\ regions, except for bright AGN if present. 
X-ray spectra are always examined and only the soft, thermal component is included in the correlation analysis.
For CGCG~097-092 and GMP~2923, the H$\alpha$ absorption is strong in the galaxy so the continuum subtraction from the {\em Subaru} narrow-band imaging data is uncertain. We elect to exclude these two galaxies in the study of the H$\alpha$ - X-ray correlation for the galaxy regions.

\begin{figure*}[!htbp]
\begin{center}
\includegraphics[angle=0,width=1.1\columnwidth]{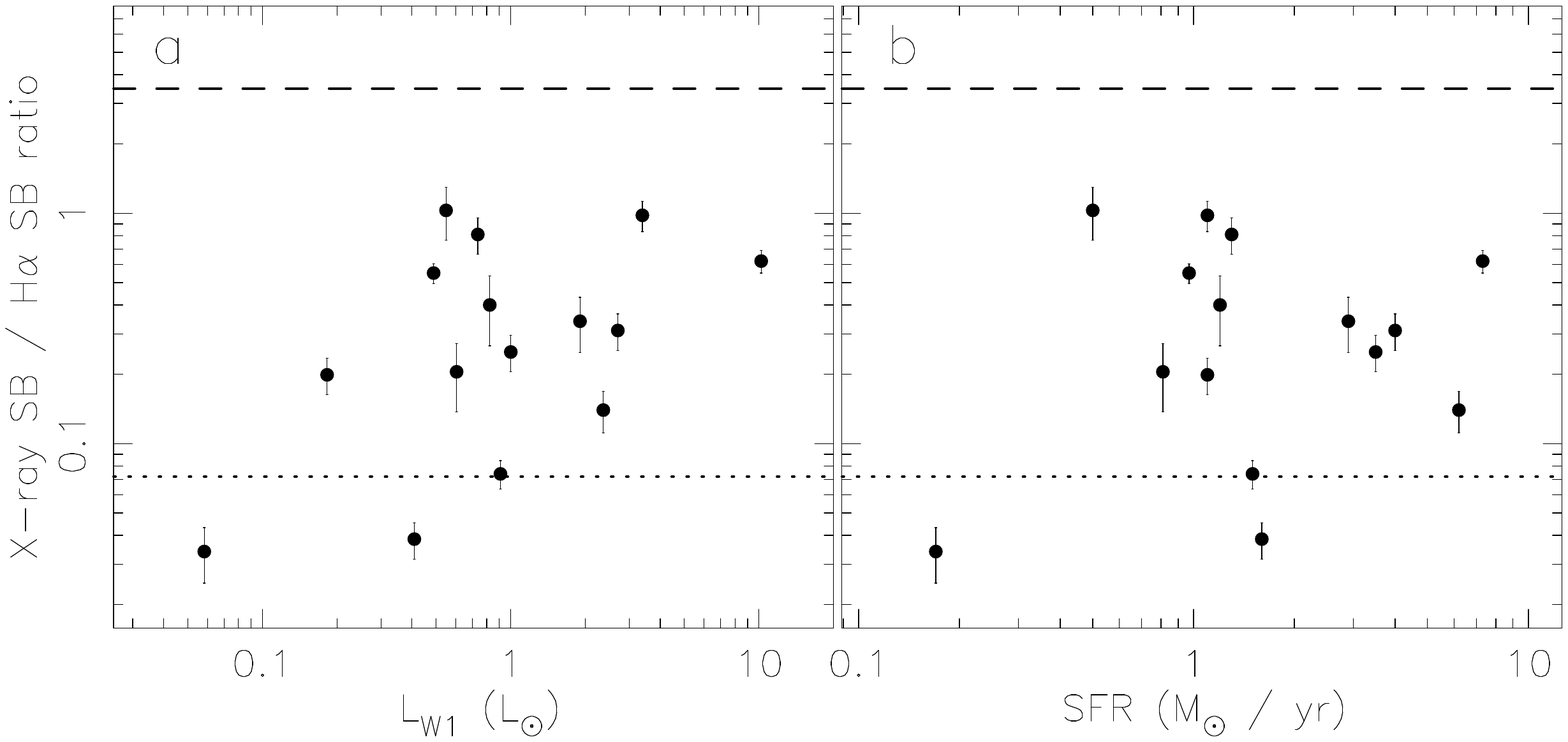}
\vspace{-1cm}
\caption{\textbf{The SB$_{\rm X}$ / SB$_{\rm H\alpha}$ ratio of the galaxy region vs. the galaxy property for 15 galaxies in our sample.} (a) the {\em WISE} W1 luminosity, (b) the SFR.
Errors are 1-$\sigma$.
The dashed line shows the average ratio in the tail regions derived from this work. The dotted line shows the ratio of 0.072, derived from the SFR - X-ray relation from \cite{2009MNRAS.399..487R} as discussed before.
}
\label{fig:relation2g2}
\end{center}
\end{figure*}

The X-ray / H$\alpha$ ratios of the galaxy regions are plotted with the {\em WISE} W1 luminosity (as a proxy of the stellar mass) and the SFR of the galaxy, as shown in Fig.~\ref{fig:relation2g2}. For \ga\ and \gb, all galaxy regions are combined. The X-ray / H$\alpha$ ratios in the galaxy regions, with a large scatter, are on average $\sim$ 10 times smaller than those in the tail regions. 
In the galaxy regions, the inclusion of the \hii\ regions boosts the H$\alpha$ luminosity while there is little soft and thermal X-ray enhancement associated with \hii\ regions, as discussed previously. The low X-ray / H$\alpha$ ratios in the galaxy regions likely reflect the much stronger contribution of SF in the galaxy regions than in the tail regions. For \ga\ and \gb\ with multiple galaxy regions, one can also see the gradual increase of the X-ray-to-H$\alpha$ ratio from the nuclear region to the tail regions.

We can also use the results on the galaxy regions to understand the impact of \hii{} regions or SF on the X-ray-to-H$\alpha$ ratio in tails. If we assume the X-ray-to-H$\alpha$ ratio is 3.5 in tails without any SF and the ratio is 10 times smaller in the galaxy, the decrease on the ratio would be 0.89$a$, if the contamination of the SF in the tail is $a$. Thus, a 10\% contamination would translate to 8.9\% decrease on the ratio and a 50\% contamination would translate to 44.5\% decrease on the ratio.
More detailed correlation analysis in the galaxy regions (to a less extent also in the tail regions) requires better separation of different emission components in H$\alpha$ with sensitive IFS data, which is beyond the scope of this work.

\subsection*{The X-ray/H$\alpha$ ratios with the distance to the galaxy}

We also examined whether the X-ray/H$\alpha$ ratios change with distance from the galaxy as continuing mixing between the stripped ISM and the surrounding ICM should eventually deplete the cold gas. This can be done for seven galaxies with more than one tail region studied, as shown in Supplementary Fig.~\ref{fig:ratio_dist}.
While an increase of the ratio at large distances from the galaxy is suggested by the data of \ga, there is not any significant evidence for the ratio changing within $\sim$ 60 kpc from the nucleus for the other six galaxies. On the other hand, this ratio is likely elevated for isolated clouds that have been separated from the parent galaxy for a long time (e.g., \cite{Ge21}).

\begin{figure*}[t]
\begin{center}
\vspace{-0.5cm}
\includegraphics[angle=0,width=1.0\columnwidth]{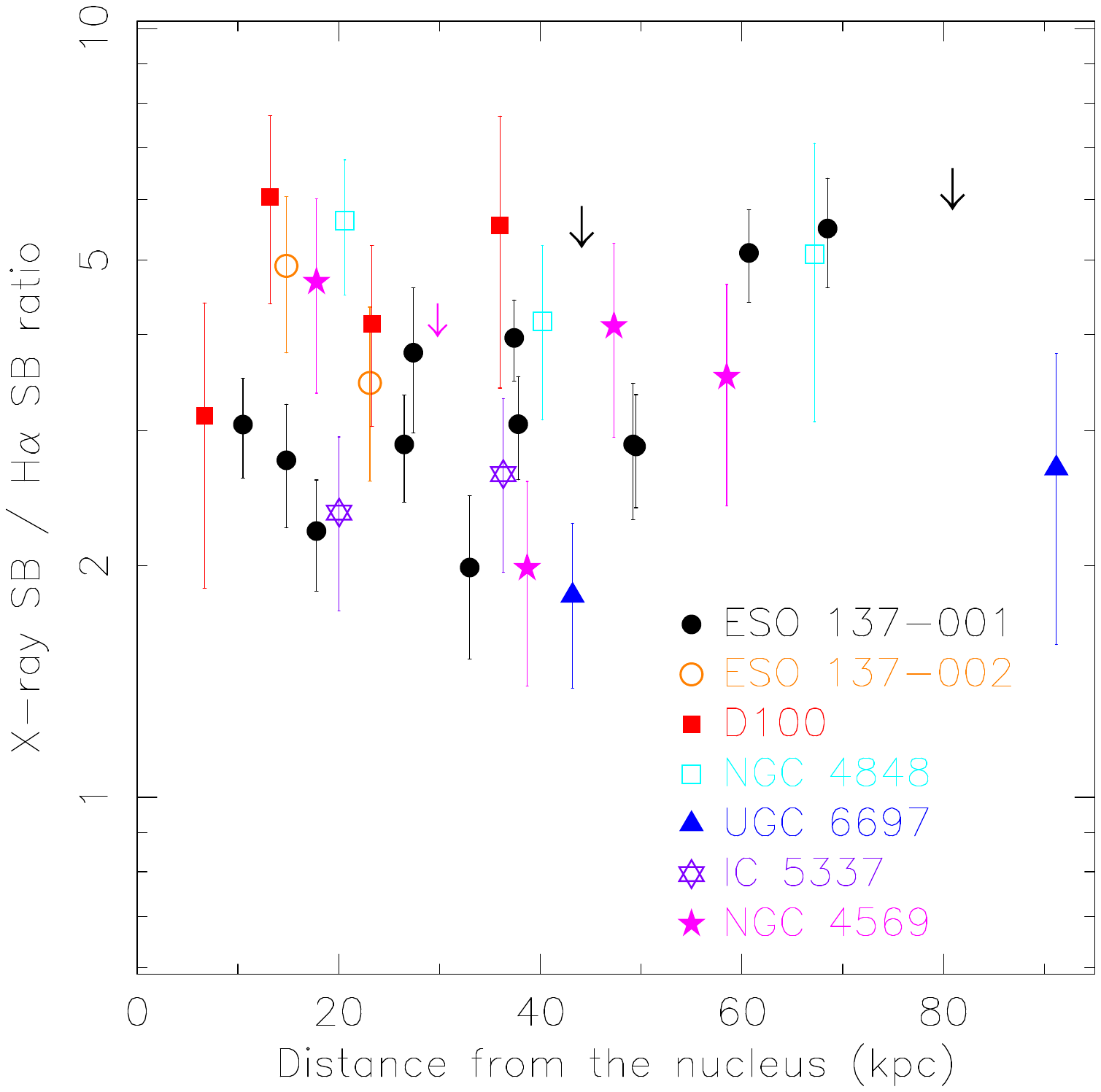}
\vspace{-1.1cm}
\caption{\textbf{The X-ray/H$\alpha$ SB ratio vs. distance to the galactic nucleus for seven galaxies with more than one tail region}. Errors for detections are 1-$\sigma$. Upper limits are 5-$\sigma$.
}
\label{fig:ratio_dist}
\end{center}
\end{figure*}

\subsection*{The implications of the H$\alpha$ - X-ray correlation}

Here we present a simple model to study the H$\alpha$ - X-ray correlation discovered. For a fixed surface area in the tail and assuming uniform densities for both the hot and warm gas, we can write the total X-ray luminosity and the total H$\alpha$ luminosity in the area as:

\begin{equation}
L_{\rm X} = n_{\rm e, X} n_{\rm p, X} \Lambda(T, Z) V f_{\rm X}
\end{equation}

\begin{equation}
L_{\rm H\alpha} = n_{\rm e, H\alpha} n_{\rm p, H\alpha} \alpha_{\rm H\alpha} h \nu_{\rm H\alpha} V f_{\rm H\alpha}
\end{equation}

where $n_{\rm e, X}$ and $n_{\rm p, X}$ are the electron and proton number densities in the hot, X-ray emitting phase; $n_{\rm e, H\alpha}$ and $n_{\rm p, H\alpha}$ are the electron and proton number densities in the warm, H$\alpha$ emitting phase; $\Lambda$ is the cooling function that depends on temperature ($T$) and abundance ($Z$); $\alpha_{\rm H\alpha}$ is the recombination coefficient for the H$\alpha$ emission; $h \nu_{\rm H\alpha}$ is the energy of an H$\alpha$ photon; $V$ is the volume projected on this surface area;
$f_{\rm X}$ and $f_{\rm H\alpha}$ are the volume filling factors for the hot and warm gas in the volume $V$. We can also simply assume that the hot gas and the warm gas have the same mean molecular weight.

Assuming a typical X-ray cooling function of 10$^{-22.5}$ erg s$^{-1}$ cm$^{3}$ (e.g., \cite{2009A&A...508..751S}) for the $T \sim 10^{7}$ K gas in the tail and a case-B recombination coefficient of 1.17$\times10^{-13}$ s$^{-1}$ cm$^{3}$ for $T = 10^{4}$ K warm, ionized gas (\cite{2006agna.book.....O}), we have:

\begin{equation}
\frac{L_{\rm X}}{L_{\rm H\alpha}} \approx 89 (\frac{n_{\rm e, X}}{n_{\rm e, H\alpha}})^2 \frac{f_{\rm X}}{f_{\rm H\alpha}}
\end{equation}

Our work shows the ratio is $\sim$ 3.5. If we assume that the hot gas and the warm gas are in pressure equilibrium, $\frac{n_{\rm e, X}}{n_{\rm e, H\alpha}} = \frac{T_{\rm H\alpha}}{T_{\rm e, X}} \sim 10^{-3}$, we can derive $\frac{f_{\rm X}}{f_{\rm H\alpha}} \sim 4\times10^{4}$. If the X-ray gas is volume filling with $f_{\rm X}$ close to 1 (e.g., \cite{Sun10s,Zhang13s}), the implied filling factor for the warm, ionized gas is very small, $\sim 2.5\times10^{-5}$. This is about two orders of magnitude smaller than what was assumed in some previous works (e.g., \cite{Sun10s,Zhang13s}), although the actual value of $f_{\rm H\alpha}$ is poorly known. While one may want to relate $f_{\rm H\alpha}$ to the observed covered fraction of H$\alpha$ (typically 5\% to close to 100\% in selected regions, see Supplementary Fig.~\ref{fig:eso137001b} to \ref{fig:5337}), such a conversion depends on the geometry of the H$\alpha$ clouds. If we assume that H$\alpha$ emission originates from thin sheets between the cold stripped ISM and the hot gas, the covered fraction is about $f_{\rm H\alpha}$ $V^{1/3}$ / $w$, where $w$ is the width of the sheet. With a region size of 10 kpc, a covered fraction of 10\% and the estimated $f_{\rm H\alpha}$ above, $w \sim 2.5$ pc. If instead H$\alpha$ emission comes from numerous small clouds, like sprinkles permeating the X-ray gas, a simple estimate shows that the number of small clouds is proportional to $f_{\rm H\alpha}^{-2}$ at a fixed region scale (e.g., 10 kpc). For a covered fraction of 10\% at that scale, the radius of small clouds is $\sim$ 2 pc. However, it is unclear whether the hot and warm gas are in pressure equilibrium, given the likely existence of turbulence and the continual gas mixing/evaporation. Both the hot gas and the warm gas may have a continuous temperature distribution to complicate the above estimate. The contribution from charge exchange on both X-ray and H$\alpha$ emission, also proposed for both galactic winds and X-ray cool cores (e.g., \cite{Lallement04,Zhang14}), may also be significant.
One also has to include the molecular gas in the study. Further understanding of this ratio will require more detailed studies with numerical simulations. 

\newpage
\subsection*{Comparison with galactic winds and X-ray cool cores}

We can compare the H$\alpha$ - X-ray correlation in stripped tails with potentially similar correlations in galactic winds and X-ray cool cores.

The H$\alpha$ - X-ray correlation in galactic winds beyond the galactic disk has been well known (e.g., \cite{Cecil02}, \cite{Strickland02s}). The correlation at large scales (e.g., $>$ 1 kpc) is generally good \cite{Strickland04}, although it may become worse or even become anti-correlated at smaller scales (e.g., $\sim$ 0.1 kpc for M82, \cite{Liu12}, but see \cite{Cecil02} for NGC~3079). This is not surprising as two phases should occupy different volumes at small scales. We also point out that such analysis at small spatial scales is easier for nearby starburst galaxies like M82 and NGC~253 with distances of 3.5 - 3.6 Mpc, compared with cluster galaxies in our sample with a median distance of $\sim$ 27 times larger.
X-ray emission from galactic winds is typically very soft with temperatures of 0.2 - 0.4 keV (\cite{Strickland02s},\cite{Strickland04}), lower than those for stripped tails (Supplementary Table~\ref{table:galaxy2}).

\cite{Strickland02s} examined the H$\alpha$ - X-ray correlation in NGC~253 with the \chandra\ data. They presented the H$\alpha$-to-X-ray ratios in six regions at several kpc scales beyond the disk as 0.44 - 2.01, with the X-ray band in the 0.3 - 2 keV band. Similar to X-ray stripped tails, the galactic winds in X-rays have multiple temperatures and the X-ray spectra are poorly fit with a single-$T$ model. One can have detailed spectral analysis on several examples of X-ray winds with deep X-ray data (e.g., \cite{Strickland02s},\cite{Strickland04}). If we simply adopt the best-fit model with a single-$T$ for NGC~253's X-ray winds, or 0.4 keV with an abundance of 0.02 solar, the bolometric correction factor is 2.48. This gives the bolometric X-ray vs. H$\alpha$ ratios of 1.2 - 5.6 for those regions in NGC~253's X-ray winds, very similar to the corresponding ratios for stripped tails. Unfortunately except for the above study, there is not a systematic sample study on this ratio at different spatial scales in galactic winds to our knowledge.

For X-ray cool cores, \cite{Fabian03s} gave the X-ray / H$\alpha$ SB ratio of one region in the X-ray cool core of the Perseus cluster as 0.18, which is $\sim$ 20 times smaller than the ratio found in stripped tails in this work. 
The detailed analysis by \cite{Sanders07} suggests a good correlation between H$\alpha$ and cool X-ray components at $\sim$ 5 kpc scales, which was also observed in M87 (e.g., \cite{Werner10}).
H$\alpha$ filaments in X-ray cool cores typically have much higher SB than the stripped H$\alpha$ tails, e.g., NGC~1275's H$\alpha$ filaments \cite{Conselice01} are $\sim$ 70 times brighter than \ga's H$\alpha$ tail. X-ray cool cores are bright so searching for soft X-ray enhancement in H$\alpha$ filaments of X-ray cool cores can be challenging.
Sample studies are required in X-ray cool cores to better understand the H$\alpha$/X-ray correlation there.

\setcounter{enumiv}{0}
\bibliographystyle{naturemag}
\bibliography{tailss.bib}

\end{document}